\documentclass[apj]{emulateapj}
\usepackage{xspace}
\usepackage{amsmath}
\usepackage{epsfig}
\usepackage{graphicx}
\usepackage{hyperref}
\usepackage{aas_macros}

  \renewenvironment{equation}{
    \begin{oldequation}
\setlength{\thinmuskip}{3mu}
\setlength{\medmuskip}{4mu}
\setlength{\thickmuskip}{5mu}
  }{ \end{oldequation} }

\def\nW{$\mathrm{nW~m^{-2}sr^{-1}}$}
\def\erg{$\mathrm{erg~s^{-1}cm^{-2}}$}
\def\ergdeg{{\rm erg~s^{-1}cm^{-2}}}
\def\mic{$\mu {\rm m}$}
\def\gsim{\gtrsim}
\def\lsim{\lesssim}

\begin{document}

\pagestyle{myheadings} \markright{DRAFT: \today\hfill}

\title{ The Contribution of \lowercase{z}$\lsim 6$ Sources to the Spatial Coherence in the Unresolved Cosmic near-Infrared and X-ray Backgrounds }

\author{ K. Helgason$^{1,2}$, N. Cappelluti$^{3,4}$, G. Hasinger$^5$, A. Kashlinsky$^{2,6}$, M. Ricotti$^1$ }
\affil{$^1$Department of Astronomy, University of Maryland, College Park, MD 20742; kari@astro.umd.edu}
\affil{$^2$Observational Cosmology Laboratory, Code 665, NASA Goddard Space Flight Center, Greenbelt MD 20771}
\affil{$^3$INAF-Osservatorio Astronomico di Bologna, Via Ranzani 1, 40127 Bologna, Italy}
\affil{$^4$University of Maryland, Baltimore County, 1000 Hilltop Circle, Baltimore, MD 21250}
\affil{$^5$Institute for Astronomy, University of Hawaii, 2680 Woodlawn Drive, Honolulu, HI 96822}
\affil{$^6$SSAI, Lanham, MD 20706 }
\email{kari@astro.umd.edu}

\begin{abstract}

A spatial clustering signal has been established in {\it Spitzer}/IRAC measurements of the unresolved Cosmic near-Infrared Background (CIB) out to large angular scales, $\sim$1 deg.  This CIB signal, while significantly exceeding the contribution from the remaining known galaxies, was further found to be coherent at a highly statistically significant level with the unresolved soft Cosmic X-ray Background (CXB). This measurement probes the unresolved CXB to very faint source levels using deep near-IR source subtraction. We study contributions from extragalactic populations at low to intermediate redshifts to the measured positive cross-power signal of the CIB fluctuations with the CXB.  We model the X-ray emission from AGN, normal galaxies and hot gas residing in virialized structures, calculating their CXB contribution including their spatial coherence with all infrared emitting counterparts. We use a halo model framework to calculate the auto and cross-power spectra of the unresolved fluctuations based on the latest constraints of the halo occupation distribution and the biasing of AGN, galaxies and diffuse emission. At small angular scales ($\lsim 1^\prime$), the 4.5\mic\ vs 0.5--2 keV coherence can be explained by shot noise from galaxies and AGN. However, at large angular scales ($\sim 10^\prime$) we find that the net contribution from the modeled populations is only able to account for $\sim$3\% of the measured CIB$\times$CXB cross-power. The discrepancy suggests that the CIB$\times$CXB signal originates from the same unknown source population producing the CIB clustering signal out to $\sim$1 deg.

 
\end{abstract}


\section{Introduction}

Background radiation is produced by a variety of sources that dominate different parts of the electromagnetic spectrum and includes contributions from sources inaccessible to individual telescopic studies. The near-IR Cosmic Infrared Background (CIB) probes the history of starlight and associated emission falling into the 1--5 \mic\ range \citep[see review by][]{Kashreview} whereas the Cosmic X-ray Background (CXB) probes emissions from accretion powered sources and hot ionized gas. Spatial correlations between the two can arise from sources emitting at both IR and X-ray wavelengths or from separate IR and X-ray sources that share the same large scale structures. Angular fluctuations in the CIB have been revealed after carefully masking resolved sources in deep near infrared (NIR) exposures and Fourier analyzing the pixels remaining in the maps. This technique has led to measurements using {\it Spitzer}/IRAC, {\it HST}/NICMOS and {\it AKARI}/IRC \citep{KAMM1,KAMM2,Thompson07a,Matsumoto11,Kashlinsky12,Cooray12b}. The amplitude of the CIB fluctuations in the {\it Spitzer} and AKARI data, now extending to $\sim 1^\circ$, is well above the expected contribution of local foregrounds such as the Zodiacal Light and Galactic cirrus and implies an unresolved extragalactic CIB component which significantly exceeds the power from the remaining known galaxy populations \citep{Helgason12}. The minimal CIB flux implied by the new sources responsible for these fluctuations is of order 0.5-2 \nW \citep{KAMM2} and is well below the claimed direct CIB flux measurements from DIRBE \citep{DwekArendt98}, IRTS \citep{Matsumoto05} and AKARI \citep{Tsumura13}, being consistent with limits placed by $\gamma$-ray attenuation from very high energy sources \citep{Meyer12,Ackermann12,HESS13}. The power spectrum of the {\it Spitzer} fluctuations is highly isotropic on the sky and is consistent with that of high-$z$ $\Lambda$CDM clustering out to $\sim$1$^\circ$ \citep{Kashlinsky12}; it appears, within the statistical uncertainties, with the same amplitude and spatial shape at both deeper (lower shot-noise) \citep{KAMM3} and shallower (larger shot-noise) \citep{Cooray12b} IRAC maps.

The measured spatial power spectrum of the source-subtracted CIB fluctuations rises at angular scales $\gsim 20^{\prime\prime}$. The amplitude and shape of this rise is a direct measurement of the clustering properties of the underlying source populations, and thus a primary key or understanding their nature. The interpretation of these power spectra has been a matter of debate \citep{Salvaterra06,KAMM3,Fernandez10,Cooray12a,Yue13a}. It is now firmly established that the extragalactic signal is inconsistent with the emission of presently resolved galaxies ($z<6$) with extrapolation to low luminosities \citep{Helgason12}. On the other hand, it was proposed that CIB fluctuations from the first light era could be measurable thus making the signal a critical tool for studying the high-$z$ Universe \citep{Kashlinsky04,Cooray04}. The CIB fluctuations show no spatial correlations with deep Hubble/ACS maps $\lsim$0.9\mic\ implying that the sources are extremely faint and/or that their Lyman-break is redshifted well into the near-IR, $(1+z)0.1$\mic\ $\gsim$0.9\mic\ \citep{KAMM4}. Current measurements show strong clustering on large scales coupled with the low shot noise levels on small scales \citep{KAMM3}. This is consistent with a high redshift origin although an alternative scenario recently has been proposed to explain some of these measurements as arising from the intrahalo light from stars stripped of their paternal halos at intermediate redshifts of $z\sim (1-4)$ \citep{Cooray12b}  (see Sec. \ref{sec_ihl} for full discussion).

The large scale clustering of the source-subtracted CIB fluctuations does not provide direct information on whether the underlying sources are powered by stellar nucleosynthesis or accretion onto compact objects. \citet{Cappelluti13} (C13, hereafter) provided observational evidence for a substantial population of accreting sources among the CIB sources raising the intriguing possibility of extensive black hole activity in the early Universe. C13 used deep source-subtracted {\it Spitzer}/IRAC and Chandra maps of a common region to reveal a highly significant cross-power (3.8,5.6)$\sigma$ between the unresolved CIB at 3.6,4.5\mic\ and the soft 0.5-2 keV CXB arguing for a high-$z$ origin of the sources. An interesting and specific model for the discovered CIB$\times$CXB signal explained it in terms of direct collapse black holes at $z\gsim 12-15$ \citep{Yue13b} by a mechanism which is capable of reproducing both the unexplained CIB and cross CIB$\times$CXB fluctuation signals without violating constraints imposed by the total measured soft CXB. However, before models of such hypothetical high-$z$ sources can be favored as leading explanations for the measured signal, a more quantitative analysis of known, and proposed, source classes at $z\lsim 6$ is needed. There exists a variety of known mechanisms of X-ray production which can spatially correlate with optical/IR emitting counterparts in complex ways.

The deepest Chandra surveys have been able to resolve $\sim 80-90\%$ of the [0.5-8] keV CXB into individual point sources, the majority of which is made up of AGN \citep{HickoxMarkevitch06,Lehmer12}. However, at the faintest fluxes, the abundance of sources identified as normal galaxies rapidly approaches that of AGN and is likely to dominate at fainter levels. The X-ray emission within galaxies comes predominantly from X-ray binaries (XRBs), a compact object accreting from a companion star, which have been found to scale well with galaxy properties such as star formation rate and stellar mass \citep[e.g.][]{Ranalli03,PersicRephaeli07,Lehmer10}. These sources have recently been detected out to deeper levels, and higher redshifts, by stacking analyses \citep{Cowie12,Basu-Zych13}. \citet{Cappelluti12} studied the unresolved CXB fluctuations ($\gsim 2\times 10^{-16}$\erg) remaining in deep Chandra exposures and determined that the bulk is produced by gas residing in galaxy groups and clusters ($\sim$50\%), with the rest being contributed by AGN and galaxies. At these levels, any contribution from high-$z$ miniquasars would be overwhelmed by these low-$z$ components ($<$5\%), although the systematic uncertainty in the mean level of the CXB increases such constraints by a large factor. The signal measured in C13 is revealed only after eliminating undetected X-ray sources down to unprecedented flux levels, $\ll 5\times 10^{-17}$\erg, using the deep source-subtracted {\it Spitzer} maps.

In this paper, we explore the contribution of the intermediate $z$ sources to the measured level of the  CIB$\times$CXB coherence (C13) by modeling components from different populations: galaxies, AGN and diffuse emission, in both IR and X-rays. We also discuss the intrahalo model of \citet{Cooray12b} vs the full set of CIB constraints and its measured coherence with the unresolved CXB. This exhausts the set of known and proposed populations out to $z\lsim 4-6$. We present a formalism for reconstructing the cross power spectrum of the fluctuations produced by each source class using the latest observational evidence for their clustering and abundance. 

The paper is organized as follows: following Section \ref{sec_def} which defines the basic parameters, Section \ref{sec_c13} discusses the CIB$\times$CXB measurement of C13 in more detail. Section \ref{sec_cib} addresses contributions to the CIB and presents population modeling of optical/IR emitting sources. In Section \ref{sec_cxb} we do the same for CXB contributions with X-ray population models. In Section \ref{sec_cross} we develop the formalism for reconstructing the CIB$\times$CXB fluctuation signal and present our results in Section \ref{sec_results}, followed by discussion in Section \ref{sec_discussion}.

\section{Definitions and parameters} 
\label{sec_def}

\begin{table*}
\centering
\caption{Technical definitions }

\begin{tabular}{ l c c c }
\hline\hline
Name & Symbol & Expression & Units$^a$ \\
 & & & \\
\hline
Infrared flux$^b$ & $F$ & $\nu I_\nu$ & \nW\ \\
X-ray flux$^c$    & $S$ & $\int_{E_1}^{E_2}N(E)dE$ & ${\rm erg~ s^{-1}cm^{-2}sr^{-1}}$ \\
Angular scale & $\theta$ & $2\pi/q$ & arcsec ($^{\prime\prime}$) \\
Angular wavenumber & $q$ & $kd_c$ & rad$^{-1}$ \\
Sky brightness   & $F(\theta)$ & $\langle F \rangle + \delta_\theta$ & \nW\ \\
Two dimensional Fourier transform & $\Delta_q$ & $\int \delta_\theta e^{-iq\cdot \theta}d\theta$ & ${\rm nW~ m^{-2}rad^{-1}}$ \\
Angular power spectrum & $P(q)$ & $\langle \Delta \Delta^* \rangle$ & ${\rm nW^2m^{-4}sr^{-1}}$ \\
Cross-power spectrum & $P_{mn}(q)$ & $\langle \Delta_m \Delta_n^* \rangle$ & ${\rm erg~ s^{-1}cm^{-2}nW m^{-2}sr^{-1}}$ \\
Fluctuations & $\langle \delta F_\theta^2 \rangle$ & $q^2P(q)/2\pi $ &  ${\rm nW^2m^{-4}sr^{-2}}$ \\
Broad band averaged power & $\langle P \rangle$ & $\int_{q_1}^{q_2}P(q)qdq/\int_{q_1}^{q_2}qdq$ & ${\rm erg~ s^{-1}cm^{-2}nW m^{-2}sr^{-1}}$ \\
Coherence & $\mathcal{C}(q)$ & $P_{mn}^2(q)/(P_m(q)P_n(q))$ & \\
 & & & \\
\hline
\end{tabular}
\tablecomments{$^a$these are the units we use throughout the paper $^b$refers to {\it Spitzer}/IRAC 3.6\mic\ and 4.5\mic\ flux. $^c$refers to soft X-ray 0.5-2 keV band flux. 
}
\label{tab_definitions}

\end{table*}
Throughout, all quantities referred to as ``X-ray''  or denoted by ``X'' correspond to the emission in the soft X-ray band 0.5-2 keV, unless noted otherwise. The suffix IR refers to the near-IR wavelengths 3.6\mic\ and 4.5\mic\ i.e. the effective wavelengths of {\it Spitzer}/IRAC bandpasses 1 and 2 whereas FIR refers to the total infrared quantities, integrated over 10--1000\mic. We assume standard $\Lambda$CDM cosmology using the cosmological parameters from the Planck experiment \citep{PlanckCosmology}. All magnitudes are in the AB system. 

Background fluctuations are characterized by the spatial power spectrum as a function of the spatial frequency $q$ (or spatial scale $2\pi/q$), defined as $P(q)=\langle |\Delta(\vec{q})|^2\rangle$, where $\Delta(\vec{q})$ is the 2-D Fourier transform of the source-subtracted CIB. The mean square fluctuation in CIB flux on angular scale $\theta=2\pi/q$ is defined as $\langle \delta F^2 \rangle \equiv q^2P(q)/(2\pi)$. The cross-power describing the correlations between fluctuations at different wavelengths $(m,n)$ is $P_{mn} (q) = \langle \Delta_{m}(q) \Delta^*_{n}(q)\rangle = {\cal R}_{m}(q) {\cal R}_{n}(q) + {\cal I}_{m}(q) {\cal I}_{n}(q)$ with ${\cal R, I}$ standing for the real, imaginary parts of the Fourier transform, $\Delta(\vec{q})$.  The cross-power spectrum is a real quantity which can assume positive or negative values.  Coherence is then defined in its usual way as in \citet{Kashlinsky12}, ${\cal C}(q) \equiv \frac{[P_{mn}(q)]^2}{P_m(q) P_n(q)}$. In the absence of common (coherent) populations at wavelengths $m$ and $n$ the cross-power, measured from a map of $N_{\rm pix}$ pixels, will oscillate around zero with a random statistically uncertainty of order $[P_m P_n]/\sqrt{N_{\rm pix}}$. Table \ref{tab_definitions} summarizes the definitions of the quantities used throughout the paper.

The source-subtracted CIB fluctuations have two components: 1) large scales $>(20-30)''$ are dominated by the clustering of an unresolved population of unknown sources, while 2) small scales are dominated by the shot(white) noise\footnote{This refers to the net white noise component, which includes the counting noise and, in some models, also the 1-halo component.}, which arises mainly from unresolved galaxy populations. The two components may arise from different populations of unresolved sources.

\section{The Measured CIB$\times$CXB Coherence} \label{sec_c13}

The spatial coherence measured between the source-subtracted CIB fluctuations and the unresolved CXB used data from the deep Chandra ACIS-I AEGIS-XD survey and The {\it Spitzer} Extended Deep Survey (SEDS) in the EGS field, where the two datasets overlap in a $\simeq 8^\prime \times 45^\prime$ region of the sky \citep[for details, see][]{Cappelluti13}. The measured cross-power between the source-subtracted IRAC maps at (3.6,4.5)\mic\ and Chandra [0.5-2] keV maps was detected, at angular scales 10$^{\prime\prime}$--1000$^{\prime\prime}$, with an overall significance of $\simeq$(3.8,5.6)$\sigma$ respectively. At the same time, no significant signal was detected between the IRAC source-subtracted maps and the harder Chandra bands. The measured coherence signal has been detected after jointly masking resolved sources down to $m_{\rm AB}$$\simeq$25 and $7\times 10^{-17}$\erg\ in IR and X-rays respectively. The signal is characterized by the cross-power spectrum, $P_{\rm IR,X}(q)$ shown in Figure \ref{fig_cross}, and exhibits a broad band averaged cross-power in the 10--1000$^{\prime\prime}$ angular range (see definitions in Table \ref{tab_definitions})
\begin{align*}
 \langle P_{\rm 3.6\mu m,0.5-2keV} \rangle  \hspace{3pt} =\hspace{3pt} 6.4 \hspace{3pt}\pm  \hspace{3pt} 1.7, \\
 \langle P_{\rm 4.5\mu m,0.5-2keV} \rangle  \hspace{3pt}= \hspace{3pt} 7.3 \hspace{3pt}\pm  \hspace{3pt} 1.3, 
\end{align*}
in units of $10^{-20} {\rm erg~ s^{-1}cm^{-2}nW m^{-2}sr^{-1}}$. We refer to the opposite ends of the measured angular interval [10$^{\prime\prime}$,1000$^{\prime\prime}$] as small and large angular scales respectively.

In this paper we examine the contribution to this signal from extragalactic populations at $z\lsim 6$ with the putative high-$z$ populations deferred to a forthcoming paper. We decompose the total power spectrum of the fluctuations into the sum of power from sources of types known to emit both in X-rays and optical/IR
\begin{equation}
  P_{\rm total} = P_{\rm galaxies} + P_{\rm AGN} + P_{\rm diffuse}
\end{equation}
with each of these components contributing both in terms of their large scale clustering on the sky as well as shot noise dominating small scale power. We refer to sources of X-ray emission arising collectively from stars, stellar remnants and gas within galaxies as ``normal galaxies'' whereas the term ``AGN'' is used in its broadest sense referring to any black hole activity in the centers of galaxies regardless of subclasses. We also consider non-point sources such as hot diffuse gas and dispersed light around galaxies which we collectively refer to as ``diffuse'' components.

The measured coherence can be interpreted as the product of the fraction of the emission due to common populations i.e. $\mathcal{C}=\zeta_m^2\zeta_n^2$, where $\zeta_m$ is the fraction contributed by the common populations in band $m$. At large angular scales ($>$20$^{\prime\prime}$) the level of the measured coherence between 4.5\mic\ and [0.5-2] keV is measured to be ${\cal C}(q) \sim 0.03-0.05$ implying that at least 15--25\% of the large scale power of the CIB fluctuations is correlated with the spatial power spectrum of the X-ray fluctuations, i.e. $\sqrt{\mathcal{C}}\gsim$15--25\%. This implies that the true nature of the signal lies somewhere in between two limiting scenarios
\begin{itemize}
\item[(1)] 100\% of the large scale CIB fluctuations are contributed by common X-ray sources, that make up $\sim$15--25\% of CXB 
fluctuations,
\item[(2)] $\sim$15--25\% of the large scale CIB fluctuations are contributed by common X-ray sources, that make up 100\% of CXB fluctuations.
\end{itemize}
Note however, that $\sqrt{\mathcal{C}(q)}$ is a scale dependent value. It is important to stress that the term ``common sources'' does not necessarily imply that the corresponding parts of the CIB and CXB are produced by the same physical sources emitting at both IR and X-rays. The IR and X-ray emitters may simply be separated by an angle smaller than the Chandra Gaussian beam of $\simeq$10$^{\prime\prime}$ corresponding to a physical scale of $\sim 0.1h^{-1}$Mpc at $z=1$. This defines the scale of the individual ``objects'' in the analysis that follows.

C13 also note that the unresolved CXB fluctuations may be contaminated by ionized gas in the Milky Way. Unfortunately, this component is diffucult to model and subtract but it is not expected to exhibit positive cross-correlation with the Galactic cirrus and should rather anti-correlate with infrared emitting dust clouds. Therefore the observed X-ray fluctuations are an {\it upper} limit for the extragalactic CXB component and its coherence with the CIB, quoted at ${\cal C}\sim (0.15-0.25)^2$, should be considered a {\it lower} limit.

\section{ Sources of the CIB } \label{sec_cib}

The CIB levels from undetected populations implied by the source-subtracted fluctuations require $\gsim$0.5 \nW\ \citep{KAMM3} on top of the extrapolated flux from known galaxies \citep{Keenan10a,Ashby13}. This level of CIB is therefore easily accommodated by both direct and indirect measurements but the sources of this component have not been conclusively identified. However, valuable insight can be obtained from population studies and deep observations at other wavelengths.

\subsection{Galaxies} \label{sec_irgal}

The extrapolation of faint galaxy populations suggests an unresolved CIB of $\sim$0.1--0.3 \nW\ at 3.6\mic\ which is mostly produced 
in the $1<z<4$ range \citep{Helgason12}. The large scale clustering of these populations, $\delta F/F \lsim 0.05$, was found to be 
insufficient to account for the observed CIB fluctuations at large scales (see Figure \ref{fig_irauto}). However, provided that unresolved 
galaxies dominate the unresolved CXB fluctuations, they could in principle produce enough coherence with the CXB to reach $
\mathcal{C}\sim 0.02-0.05$ while remaining a underdominant component in the CIB fluctuations. In this paper, we use the empirical calculation of \citet{Helgason12} in which the CIB from galaxy populations was reconstructed using over 230 observed 
multiwavelength luminosity functions measured at z$\lsim$5. The faint-end regime of this original reconstruction was extrapolated to 
low luminosities and its accuracy has since been verified in deep near-IR source counts from the SEDS survey reaching 26 AB mag 
\citet{Ashby13}. We address the X-ray emission from these unresolved galaxies in Section \ref{sec_xgal}.

\subsection{AGN}

\begin{figure}
      \includegraphics[width=0.45\textwidth]{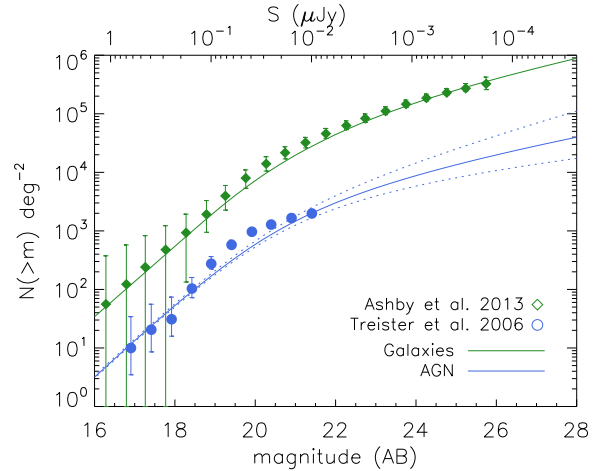}
      \caption{ The cumulative source number counts of galaxies (green line) compared with AGN (blue line). The models for AGN are constructed from J-band LFs and are uncorrected for host galaxy contribution (explained in the text). Corresponding AGN counts data is from \citet{Treister06} (circles). The dotted lines show the cases in which the faint-end slope of the AGN LF is extrapolated with a constant slope of $-$1.6 (upper) and $-$1.2 (lower). The green line shows the galaxy reconstruction of \citet{Helgason12} compared with SEDS data \citep{Ashby13} (diamonds). }
    \label{fig_ircounts}
\end{figure}

The AGN fraction of the resolved CIB sources is small, $\lsim$8--10\% \citep{Treister06}, but their leading role for the CXB makes them important for the interpretation of the CIB$\times$CXB correlation. Here we estimate the CIB production from AGN by constructing an AGN population model from J-band luminosity functions of \citet{Assef11}. This sample consists of 1838 AGN (Type 1 and 2) at 0$<$z$\lsim$6 selected in both IR and X-ray, and is therefore less affected by incompleteness and biases seen in purely optical or infrared selected samples. The choice of rest-frame J-band (1.25\mic) luminosity functions minimizes the uncertainty in the k-correction as distant populations observed at 3.6--4.5\mic\ emit light at rest-frames 0.7$\lsim$$\lambda$$\lsim$4.5\mic\ for $z<4$. We use the pure luminosity evolution parameterization of the AGN LF given by \citet{Assef11} which includes both host galaxy emission and reddening, extrapolating the parameterized evolution beyond z$>$5 (assuming LDDE model instead does not affect our results). The inclusion of the host galaxy is important since the X-ray faint AGN tend to be optically obscured and dominated by their host galaxy light which also contributes to the CIB. At every distance the 1.25\mic\ emission is corrected to $\lambda_{\rm em}=\lambda_{\rm obs}$/(1+z) using the average low resolution AGN spectral template of \citet{Assef10} where $\lambda_{\rm obs}$ is either 3.6 and 4.5\mic\ for this paper \citep[see also ][]{Richards06}. Figure \ref{fig_ircounts} shows a reasonable agreement between this projected population and {\it Spitzer}/IRAC AGN counts of \citet{Treister06}. Inaccuracies in the shape of the counts may arise from the fact that our mean AGN spectrum is corrected for host galaxy contribution and reddening.

The typical faint-end slope measured for AGN luminosity functions ranges from $-1.6$ to $-1.3$ showing marginal evidence for flattening at higher redshifts \citep{Croom09,Glikman11}. In our model, we conservatively extrapolate a non-evolving faint-end slope of $\beta =-1.4$ to account for unresolved AGN but this likely {\it overestimates} the AGN contribution for several reasons. First, because we use LFs uncorrected for host galaxy contribution which bias the faint-end slope and increase the number of faint objects \citep{Hopkins07}. This effect can be particularly pronounced in the near-IR, as the ratio of host to AGN in unobscured objects has a typical maximum at 1.6\mic. Second, the 0.5-2 keV vs 3.6\mic\ flux ratio turns towards X/O$<$0 at faint fluxes suggesting a decreasing importance of AGN contribution \citep{Civano12}. This is consistent with significant number of optically normal galaxies that are seen as hosts of low-luminosity X-ray AGN, \citep[e.g.][]{Barger05}. Diminishing nuclear activity makes the host galaxy dominate at low luminosities and the distinction between galaxy and AGN becomes less meaningful. Then these sources should be largely accounted for by our treatment of IR galaxies (see previous subsection). In addition, our treatment of normal galaxies (see Section \ref{sec_irgal}) is based on a compilation of observed galaxy LFs which do not typically exclude AGN.

From Figure \ref{fig_ircounts}, it is clear that normal galaxies are far more numerous than AGN at all flux limits regardless of the extrapolation to faint levels. The CIB contribution of AGN is always $\lsim 10\%$ of the galaxy contribution and their sky surface density is $\sim$2--3 orders of magnitude smaller than the $\sim$1 arcsec$^{-2}$ required to explain the measured CIB fluctuations \citep{KAMM3}. AGN would need to live in low mass halos to exist in sufficient numbers at $z\lsim 4$ and our mask eliminates most halos $\gsim 10^{12}M_\odot$ where AGN are typically found. Furthermore, a recognizable signature of faint AGN is that their near-IR spectrum should increase with wavelength due to their dusty torus emission but this is inconsistent with the blue colors of the 
measured CIB fluctuations \citep{Matsumoto11}. We therefore assert that the IR emission from AGN themselves is insufficient to produce significant unresolved CIB fluctuations. However, this does not necessarily eliminate AGN as sources of the CIB$\times$CXB correlation as their X-ray emission can produce stronger correlation with other IR populations sharing common large scale structures (see Section \ref{sec_results}). 

\subsection{Diffuse emission, intrahalo light}
\label{sec_ihl}

Recently, some modeling of the origin of the CIB fluctuations has focused on a form of ``missing light'' associated with galaxy 
populations but distributed in diffuse structures around masked sources. There are several empirical lines of observational evidence 
that argue strongly against such an origin:
\begin{enumerate}
\item There are no spatial correlations between the source-subtracted {\it Spitzer}/IRAC maps and galaxies detected in deep Hubble/ACS maps (0.5-0.9\mic)  $\lsim$28 mag \citep{KAMM4}. However, there are very significant correlations between the ACS galaxies and the unmasked {\it Spitzer} maps. This means that ACS galaxies and any associated diffuse emission, cannot contribute significantly to the large scale CIB fluctuations found in source-subtracted {\it Spitzer} data.
\item  \citet{Arendt10} show that the large scale CIB fluctuations are not sensitive to increasing/decreasing the size of masked regions 
around resolved galaxies. Indeed, Fig. 17 of \citet{Arendt10} shows that there is little variation in the CIB fluctuation as the source 
masking is eroded or dilated to masking fractions varying from $\simeq 7\%$ to $\simeq 46\%$.
\item \cite{KAMM1} and \citet{Arendt10} show that there are no correlations between the source-subtracted CIB fluctuations and
the identified removed extended sources. Moreover,  \citet{Arendt10} constructed artificial halos around masked sources and 
demonstrated that the diffuse emission in the final image does not correlate spatially with the halos around masked sources that mimic missing light.
\end{enumerate}
These arguments contradict scenarios invoking any form of ``missing light'' associated with masked galaxies.

One such scenario was proposed by \citet{Cooray12b} who considered diffuse starlight scattered around and between galaxies at 
$z\sim 1-4$ as an alternative explanation for the origin of the unresolved CIB fluctuations.
\footnote{The empirical results 1--3 above were omitted from the discussion by \citet{Cooray12b}.}.
This can arise from stars stripped in mergers or ejected via other processes. 
In this paper, we consider such a diffuse component following definitions in \citet{Cooray12b} deriving the CIB production history 
as
\begin{equation}
  \frac{dF_{\rm IHL}}{dz} = \frac{c}{4\pi}\int_{M_{\rm min}}^{M_{\rm max}} L_{\rm IHL,\lambda^\prime}(M,z)\frac{dn}{dM} dM \frac{dt}{dz}(1+z)^{-1}
  \label{eq_ihl}
\end{equation}
where $\lambda^\prime=\lambda_{\rm obs}(1+z)$ using a spectral template of a typical elliptical galaxy containing old stellar populations from Starburst99 \citep{Leitherer99}. This template assumes a 900 Myr old stellar population forming at the same time with a Salpeter IMF (1--100$M_\odot$) and metallicity $Z=0.008$. In Figure \ref{fig_dfdz} we show that different choices of SED parameters do not affect the IHL flux to a great extent (unless the population is very young $<$20 Myr). In all other respects we follow the formulation in \citet{Cooray12b} recovering a flux of 1 \nW\ at 3.6\mic\ which is consistent with but slightly higher than the 0.75 \nW\ quoted in \citet{Cooray12b}. We are unable to get the large scale clustering up to the quoted $\delta F/F = 10-15\%$ and as a result our large scale fluctuations are a factor of $\sim$2--3 lower; this actually provides a somewhat better fit to the measured signal 4.5\mic\ (see Figure \ref{fig_irauto}).

As is shown in Figure \ref{fig_dfdz}, the CIB from this component, Eq. \ref{eq_ihl},  {\it exceeds} the {\it total} CIB from all galaxies already at $z\gsim 2$ according to the reconstruction of \citet{Helgason12}.
Although this by itself makes the model non-viable, we calculate the coherence levels with between CIB and diffuse X-ray emission for this component.

\section{Sources of the CXB} \label{sec_cxb}

\subsection{Normal X-ray Galaxies} \label{sec_xgal}

The bulk of the CXB ($\sim 80-90\%$) has been resolved into point sources, most of which is contributed by AGN \citep{HickoxMarkevitch06,Lehmer12}. However, the number counts from the Chandra deep fields reveal that the contribution of normal galaxies (mostly XRBs) approaches that of AGN at the faintest levels \citep{Xue11,Lehmer12}. A simple extrapolation of the slope of galaxies implies that they will ultimately dominate AGN at $\lsim10^{-17}$\erg\ (see Fig. \ref{fig_xcounts}); this is in fact suggested by deep stacking analyses \citep{Cowie12}. Although \citet{Helgason12} demonstrated the low contribution of known galaxy populations to the unresolved CIB, their CXB$\times$CIB amplitude ultimately depends on their X-ray properties.

The X-ray galaxy luminosity function (XLF) derived from {\it Chandra} and {\it XMM-Newton} data is limited to small samples of local galaxies of $L_X\gsim 10^{40}$erg/s \citep{Norman04,Tzanavaris08} with luminosity evolution consistent with $L^\star\propto (1+z)^{2.3}$ out to z$\sim$1. The X-ray emission in galaxies is dominated by a population of compact objects accreting from a stellar companion although hot gas can contribute substantially to the soft X-ray flux ($\lsim 1$ keV). To gain a better understanding of the X-ray galaxy population and its evolution a popular approach takes advantage of empirical correlations of X-ray luminosity with various galaxy properties derived from longer wavelength data. The XLF can be related to LFs measured at other wavelengths following \citet{AvniTananbaum86}
\begin{equation}
 \Phi(\log{L_X}) = \int_{-\infty}^\infty \Phi(\log{L_Y})P(\log{L_X}|\log{L_Y}) d\log{L_Y}
\end{equation}
(in num$\cdot$Mpc$^{-3}$dex$^{-1}$) where $Y$ represents a rest-frame band which correlates with X-ray luminosity via some specified $L_X$-$L_Y$ relation and $P(\log{L_X}|\log{L_Y})$ describes the probability of a source of $L_Y$ having an X-ray luminosity of $L_X$. Here we assume the probability distribution to be Gaussian
\begin{equation}
  P(\log{L_X}|\log{L_Y}) = \frac{1}{\sqrt{2\pi}\sigma}\exp{\left[ -\frac{(\log{L_X(L_Y)} - \log{L_X})^2}{2\sigma^2}\right]  }
\end{equation}
where $L_X(L_Y)$ is the X-ray luminosity predicted by a $L_X$-$L_Y$ relation and $\sigma$ is the standard deviation of the scatter in the measured correlation.

We consider the X-ray luminosity of normal galaxies to be the sum of contributions from HMXBs, LMXBs and hot gas
\begin{equation}
  L_X^{\rm gal} = L_X({\rm HMXB}) + L_X({\rm LMXB}) + L_X({\rm gas})
\end{equation}
Other forms of galaxy-wide X-ray emission, such as from WD binaries and supernova remnants, have been found to be at least ten times weaker so we ignore their contribution in this paper \citep{Boroson11}. A strong correlation is found between the star formation rate (SFR) and the overall X-ray emission of the galaxy which is attributed to active star forming regions producing bright short-lived ($\lsim$100 Myr) high mass X-ray binaries (HMXBs). On the other hand, long-lived ($\gsim$1 Gyr) low mass X-ray binaries (LMXBs) have been found to correlate well with the net stellar mass in galaxies. For a full census of X-ray galaxies and their different emission mechanism, it is therefore helpful to decompose the population into late(active) and early(quiescent)-type galaxies, $\Phi^{\rm tot} = \Phi^{\rm late} + \Phi^{\rm early}$. In the picture that follows, late-types can be assumed to contain HMXBs, LMXBs as well as hot gas whereas early-types only contain LMXBs and hot gas.

\subsubsection{Late-types} \label{sec_lt}

\begin{figure*}
\begin{center}
      \includegraphics[width=.80\textwidth]{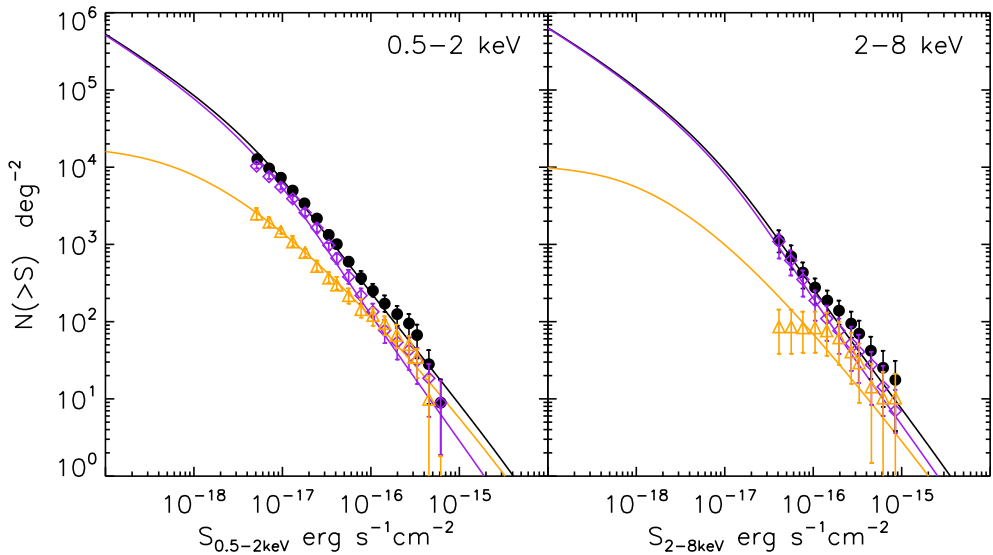}
      \caption{ Cumulative X-ray source counts in soft (left) and hard (right) X-ray bands. The panels show observed galaxy counts from the CDF-S 4Ms survey \citep{Lehmer12} where the black circles represent all galaxies and the sub-contributions from late and early-type galaxies are shown as purple diamonds and orange triangles respectively. The solid lines show the prediction of our population model in the same colors scheme (black, purple and orange for total, late and early-types respectively). We show the extrapolation all the way down to $10^{-19}$\erg\ to emphasize the expected behavior in the unresolved CXB regime. }
      \label{fig_counts}
\end{center}
\end{figure*}
The tight correlation between X-ray luminosity and SFR has established HMXBs as the dominant mechanism for X-ray production in late-type galaxies \citep[e.g.][]{Ranalli03}. For the purpose of re-constructing the evolving population of X-ray galaxies, we rely on the total infrared luminosity function (8-1000\mic)\footnote{We denote the total infrared (8-1000\mic) as ``FIR'' to avoid confusing with ``IR'' which we use to refer to the near-IR}. This choice is motivated by several points, 1) re-radiated dust emission in the FIR is an established tracer of star formation which bypasses uncertainties in UV tracers due to obscuration, 2) the $L_X$-$L_{\rm FIR}$ correlation is both very significant and well calibrated, 3) the evolution of the FIR LF has been probed out to $z\sim3$ allowing the estimation of HMXB activity for 4/5 of the cosmic time. Since the bulk CXB from galaxies comes from z$\lsim$3 it is not necessary to rely on UV LFs probing z$>$3. 
In this paper, we adopt the FIR LF measured by \citet{Magnelli09,Magnelli11} in the 0$<$z$<$2.5 range from deep {\it Spitzer} GOODS/FIDEL data. The measurement is in good agreement with other FIR LFs in the literature \citep{Takeuchi03,Rodighiero10} and is described by a double power-law parameterized by the characteristic luminosity $\L^\star$, normalization $\phi^\star$, and bright and faint-end slopes $\alpha$ and $\beta$. The evolution is consistent with pure luminosity evolution $L^\star \propto (1+z)^{3.6}$ put to $z=1$ with very mild evolution of both $L^\star$ and $\phi^\star$ at $z>1$.  Beyond $z=3$ we assume $L^\star$ gets fainter with $z$ at the rate implied by the UV LF of \citet{Bouwens11} but this has little impact on our results.

\citet{Lehmer10} study the $L_X$-SFR relation for a population of star forming galaxies covering roughly four orders of magnitude in SFR, $-1.5\lsim \log{SFR} \lsim 2.5$ in $M_\odot/{\rm yr}$. The relation is derived at 2-10 keV where hot gas becomes negligible and the emission is predominantly driven by HMXBs and LMXBs. They find a local best fit to be
\begin{equation} \label{eqn_lx2lir}
  \log{L_{2-10 keV}} = 
    \begin{cases}
      0.94\log{L_{\rm FIR}} + 30.17, &  \log{L_{\rm FIR}}\lsim 9.6 \\
      0.74\log{L_{\rm FIR}} + 32.09, &  \log{L_{\rm FIR}}\gsim 9.6
    \end{cases}
\end{equation}
with a 1$\sigma$ scatter of roughly 0.4 dex\footnote{We have replaced SFR with $L_{\rm FIR}$ (in solar units) based on the original relation used by \citet{Lehmer10}, $SFR = 9.8\times 10^{-11}L_{\rm FIR}$.}. It is important to note that because \citet{Lehmer10} derive the SFR solely based on IR luminosity our model does not depend on a $L_{\rm FIR}$-SFR calibration. The duality of the relationship in Equation \ref{eqn_lx2lir} over a wide range of SFR (0.01-100 $M_\odot$/yr) arises because LMXBs provide a non-negligible contribution to the low SFR regime. Indeed, \citet{Lehmer10} use the K-band luminosity to trace stellar mass and find a relationship $M^\star\propto$SFR$^{1.1}$ for SFR$\lsim$5$M_\odot$/yr and $M^\star\propto$SFR$^{0.3}$ for SFR$\gsim$5$M_\odot$/yr which combines SFR and $M^\star$ in a more physically justified relation $L_X$=$\alpha M^\star$+$\beta$SFR. Our adopted relation (Eqn \ref{eqn_lx2lir}) therefore accounts for both the HMXBs and LMXBs contribution in late-type galaxies. We assume an average spectral index of $\Gamma$=1.8 to convert to the 0.5-2 keV band.

Recent evidence seems to indicate that the local $L_X$-SFR relation is not constant with redshift. \citet{Basu-Zych13} find an evolution $L_X \propto (1+z)^{0.9}$ for galaxies with SFR$\gsim$5$M_\odot$/yr. However, including this evolution in our relation (Eqn \ref{eqn_lx2lir}) slightly overproduces the faintest CDFS counts \cite{Lehmer12} which could also be traced to the evolution of our FIR LF which is somewhat steeper than some other measurements in the literature \citep{Magnelli09}. We therefore include a term of  $\log{L_X}\propto 0.5(1+z)$ in Equation \ref{eqn_lx2lir} which results in good agreement with the data. If we use the observed evolution regardless ({\it overproducing} faint X-ray galaxies) this gives a CIB$\times$CXB signal which is within a factor of 1.5 larger but our final conclusions are unchanged.

\subsubsection{Early-types} \label{sec_et}

The X-ray emission from quiescent galaxies is thought to be dominated by long-lived LMXBs leftover from earlier episodes of star 
formation and hot gas in extended halos. The X-ray luminosity of early-types is found to correlate well with K-band luminosity, the 
preferred indicator of stellar mass. For the template LF for early-type galaxies we have chosen the local K-band LF of 
\citet{Kochanek01} (2MASS) for z$<$0.05 and the evolving LF \citet{Arnouts07} from combined SWIRE-VVDS-CFHTLS data reaching 
$z=2$. These measurements separate the contribution from the early-types and late-type to the total LF. The bright(high mass)-end of 
the LF (stellar mass function) is dominated by early-types whereas late-types are much more numerous at the faint(low mass)-end. 
All the K-band LFs are well described by the Schechter function with parameters $L^\star$,$\phi^\star$ and $\alpha$ measured out to 
z=2. We fit the evolving parameters in the $0<z<2$ range using the functional forms in \citet{Helgason12} and extrapolate them 
beyond $z>2$.

\citet{Boroson11} studied the distinct components of X-ray emission in nearby early-types in great detail, resolving the individual XRBs. The total LMXB luminosity is found to correlate with K-band luminosity via the relation
\begin{equation}
  \log{L_{0.3-8keV}} = \log{L_K} + 29
\end{equation}
where $L_K$ is in $L_{\odot,K}=4.82\times 10^{32}$erg$\cdot$s$^{-1}$ and the 1$\sigma$ scatter is $\approx$0.3 dex. We convert the 
0.3-8 keV luminosity to the 0.5-2 keV band assuming a spectral index of $\Gamma$=1.8.

\subsubsection{Hot gas} \label{sec_hotgas}

It is well known that galaxies contain extended hot halos of gas heated above the virial temperature emitting in lines and thermal continuum. We find that a luminosity comparable to that of XRBs is needed from hot gas in early types to account for the number of bright 0.5-2 keV sources (Fig. \ref{fig_counts}) whereas XRBs are sufficient to explain the entire late-type population. This is consistent with the fact that hot gas in early-types is found to constitute a much greater fraction of the total $L_X$ than in late-types \citep{Anderson13}. The unresolved CXB, however, cannot contain significant contributions from hot gas for several reasons. First, groups and clusters closeby with $kT$$>$1 keV are easily detected and removed in the Chandra maps and the joint IR/X-ray mask used in C13 further eliminates galaxies residing in $\gsim$10$^{12}M_\odot$ ($\sim$0.1 keV) halos out to z$\sim$2 where high-mass systems become increasingly rare. Second, normal galaxies have characteristic temperatures of $<$1 keV where their spectrum decreases exponentially. At z$>$1, their contribution quickly redshifts out of the 0.5-2 keV band. Both of these considerations act to reduce hot gas contribution in the faint unresolved regime which is dominated by low-mass systems at increasingly high redshifts.

To demonstrate this, we use the hot gas properties predicted in semi-analytic models of \citet{Guo11} mapped onto the Millennium simulations. The hot gas mass is calculated based on the baryon content, cooling and infall rate onto the halo (see \citet{Guo11} for details). We assume a density profile $\propto (1+(r/r_0)^2)^{-3\beta /2}$ with a constant $r_0 = R_{\rm vir}/10$ and $\beta = 2/3$ \citep{Anderson13} and apply k-corrections assuming a thermal continuum spectrum $\propto T_{\rm vir}^{-1/2}\exp{(-E/kT_{\rm vir})}$. In this desciption, the observed [0.5-2 keV] counts are reproduced when all systems with masses below 10$^{13}M_\odot$ are included, which is roughly the mass scale of the the most massive galaxies. However, the hot gas counts flatten towards lower fluxes where are dominated by the HMXB component. This is shown as red dotted line in Figure \ref{fig_xcounts}. In Section \ref{sec_xgas} we consider diffuse X-ray emission in more complex environments such as groups and filaments, including the warm-hot intergalactic medium (WHIM).

\subsubsection{The unresolved CXB from galaxies}

The cumulative source density seen on the sky is
\begin{equation} \label{eqn_counts}
	N(>S) = \int \int_S^\infty \frac{d^3N}{dS^\prime dzd\Omega} dS^\prime dz
\end{equation}
where the differential is expressed in terms of the XLF as
\begin{equation} \label{eqn_diff}
	\frac{d^3N}{dSdzd\Omega} = \frac{4\pi d_L^2(z)}{K(z)} \Phi(L_X,z) \frac{dV}{dz}
\end{equation}
where $dV/dz = cd_L^2(1+z)^{-1}dt/dz$ is the comoving volume element and $K(z)$ is the $k$-correction. For a power-law spectrum, $E^{-\Gamma}$, with a spectral index $\Gamma$, the $k$-correction becomes (1+z)$^{ 2-\Gamma }$. The power-law slope of X-ray galaxy spectra has been found to lie in the 1$<$$\Gamma$$<$3 range with a mean of $\approx$1.8 which is  what we assume for the HMXB and LMXB contribution. For the hot gas component we $k$-correct using a thermal continuum spectrum, $\propto T^{-0.5}\exp (-E/kT)$.

Figure~\ref{fig_counts} shows the predicted cumulative X-ray counts of our galaxy population models (Eqn. \ref{eqn_counts}) and compares them with the observed deep counts of \citet{Lehmer12} showing both the early and late-type contribution. The bright counts rise with a near-Euclidian slope and start turning over towards the faint regime as expected from both the LF turnover and cosmic expansion. At the flux limit of today's deepest measurements, $\sim 10^{-17}$\erg, the source counts are dominated by $L_{\rm FIR}^\star$ galaxies (at the knee of the LF) gradually turning over to the faint-end regime which only becomes relevant at much fainter levels ($\lsim 10^{-19}$\erg). As long as the flux converges, the unresolved CXB fluctuations are not very sensitive to the faint-end slope of the XLF.

\begin{figure}
      \includegraphics[width=0.45\textwidth]{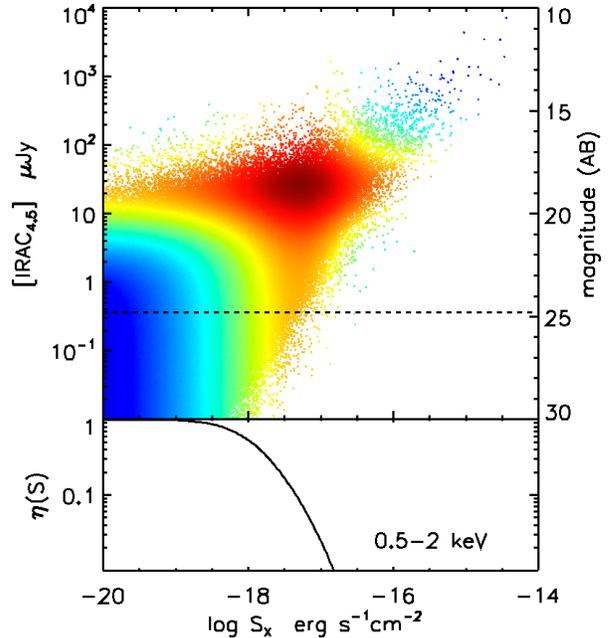}
      \caption{ {\it Top:} The 4.5\mic\ vs 0.5-2keV flux distribution of normal galaxies from the Millennium Simulations lightcones of Henriques et al. 2012. The flux limit of the {\it Spitzer}/IRAC maps of \citet{Kashlinsky12} is shown as the horizontal line. The color scheme corresponds to the log-scaled flux contribution to the total (resolved+unresolved) CIB$\times$CXB $S dN/dS$, red/blue representing large/low contribution {\it Bottom:} the unresolved selection for X-ray galaxies using an IR threshold of 25 mag extracted from this catalog (see Eqn \ref{eqn_selection}). }
    \label{fig_guo}
\end{figure}
The spatial CIB$\times$CXB correlation signal is detected by C13 after masking sources resolved in both the Chandra and {\it Spitzer}/IRAC maps. The common IR/X-ray mask leaves $\sim$68\% of the pixels in the overlapping $8^\prime \times 45^\prime$ region for the Fourier analysis. The superior resolution and depth reached in the {\it Spitzer} exposure maps causes faint X-ray sources to be subtracted well below the detection threshold of the Chandra maps. The depth of the joint CXB$\times$CIB analysis is therefore mostly determined by the IR source subtraction. In this paper, we assume a fixed magnitude limit in the {\it Spitzer}/IRAC maps with X-ray sources removed according to a selection function $\eta(S_X|m_{lim})$ where $m_{lim}$ refers to the IRAC magnitude limit of the near-IR mask. We also tried relaxing the assumption of a fixed $m_{lim}$ and instead used the complement of the source selection completeness in the SEDS EGS field. This has little effect on our results. In order to obtain $\eta(S_X|m_{lim})$ at flux levels inaccessible to current X-ray observatories, we look at the distribution of $f_X/f_{IR}$ predicted in the semi-analytic model (SAM) of \citet{Guo11} mapped onto the Millennium Simulation (for a detailed description we refer to \citet{Guo11}). Mock lightcones based on this model were constructed by \citet{Henriques12}. Although mostly consistent with galaxy counts in the optical, the semi-analytic model tends to {\it overestimate} the number of small systems causing the abundance of faint galaxies to overpredict observed 3.6\mic\ and 4.5\mic\ counts. We therefore apply a post-correction to the galaxy population by shifting excess sources of magnitude $m$ by a factor
\begin{equation}
  \Delta m = \frac{|m(n_{obs}) - m(n=n_{\rm obs})|}{\delta}
\end{equation}
where $m(n)$ are the Millennium SAM derived counts and $n_{obs}$ are the observed counts. We find that for a modification factor of $\delta=1.05$ the population of \citet{Henriques12} is brought into a good agreement with the observed counts while conserving the total number of systems and their redshift distribution. Figure~\ref{fig_guo} shows the 4.5\mic\ versus 0.5--2 keV brightness distribution of normal galaxies according to the Millennium SAM where we have used the SFR and $M^\star$ to predict the X-ray luminosity via $L_X=\alpha M^\star +\beta$SFR of \citet{Lehmer10} and include a scatter of $\sigma = 0.4$. This is essentially the same relation as the $L_X$-$L_{IR}$ in Equation \ref{eqn_lx2lir} with the conversion SFR/$L_{IR}$=$9.8\times 10^{11}$. The approximate detection limit of the {\it Spitzer} maps is shown in Figure \ref{fig_guo} as horizontal line and the color scheme scales with flux per solid angle i.e. depicting the contribution to the total CXB$\times$CIB background. From Figure \ref{fig_guo} it is clear that most of the background light is resolved and eliminated in the masking process with a diminishing contribution from the remaining unresolved sources towards the bottom left. We define the selection function of 0.5--2 keV source removal as the unresolved galaxy fraction
\begin{equation} \label{eqn_selection}
  \eta(S_X|m_{lim}) = \frac{N(S_X|m>m_{lim})}{N(S_X|m)}
\end{equation}
and display it in the bottom panel in Figure \ref{fig_guo}. This shows that for an IR limit of 25 mag, 90\% of sources are removed at $\simeq 3\times 10^{-18}$\erg\ which considerably fainter than the flux limit of CDFS \citep{Xue11}. Systems identified as subhalos in the Millennium catalog were removed together with its parent halo provided it is brighter than $m_{\rm lim}$. The subhalos have little effect on our results.

\begin{figure}
      \includegraphics[width=.47\textwidth]{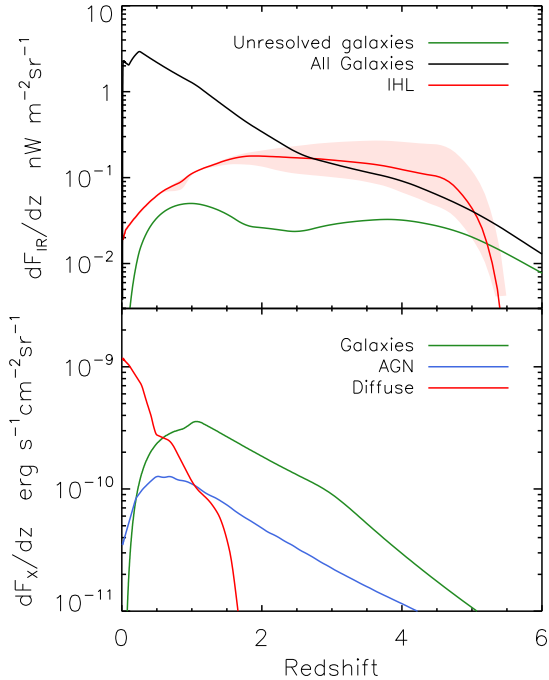}
      \caption{ {\it Top:} The unresolved CIB (4.5\mic) production rate from unresolved galaxies, all galaxies and intrahalo light (green, black ,red). The light shaded region shows the range of IHL flux for SED templates with different ages (20--900 Myr), metallicities (0.04--0.001$Z_\odot$) and IMF. Model details are explained in Section \ref{sec_cib}. The AGN contribution falls slightly below the plot range. {\it Bottom:} The unresolved 0.5-2 keV CXB production histories from galaxies, AGN and diffuse gas (green, blue, red). Models are explained in Section \ref{sec_cxb}. }
      \label{fig_dfdz}
\end{figure}
Our XLF model allows us to construct the flux production rate per solid angle from undetected galaxies as
\begin{equation}
	\frac{dF_X}{dz} = \int \eta(S) S \frac{d^3N}{dSdzd\Omega} dS
\end{equation}
where $d^3N/dS/dz/d\Omega$ comes from Equation \ref{eqn_diff} and $F_X$ is the X-ray flux per solid angle. Figure~\ref{fig_dfdz} shows the history of the emitted X-rays from galaxies remaining after removing IR sources brighter than 25 mag. Very little CXB remains at $z\lsim 0.5$ after source subtraction but rises thereafter and peaks at $z\sim 1$ close to the peak of the star formation history. 

\subsection{X-ray AGN}

\begin{figure}
      \includegraphics[width=0.45\textwidth]{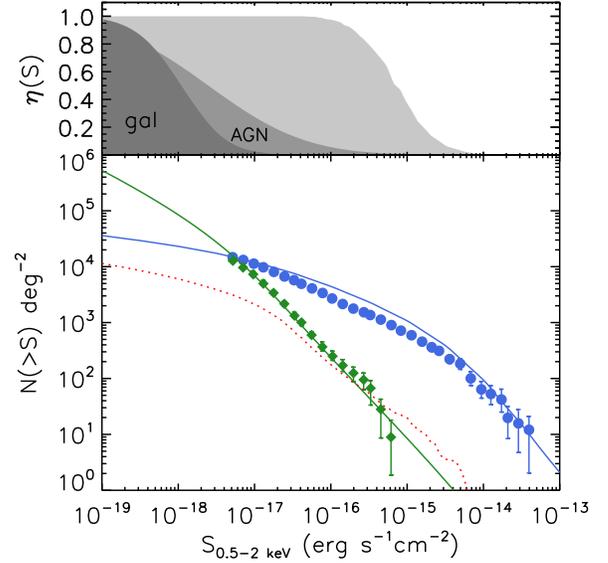}
      \caption{ {\it Upper:} The selection of the unresolved regime for galaxies (darkest gray shade) and AGN (gray shade). Note that for a constant $\rm{m_{IR}}$ limit, a larger fraction of AGN remain unresolved due to their higher X/O ratio and greater dispersion. The complement of the X-ray source selection in AEGIS-XD \citet{Goulding12} is shown for comparison (lightest gray shade). This essentially defines the unresolved regime in C13 for X-ray source removal i.e. without additional IR masking.  {\it Lower:} The source counts of \citet{Lehmer12} compared with our adopted \citet{Gilli07} model for AGN (blue), and our XRB population model (green). The dotted line shows the hot gas contribution from virialized halos. }
    \label{fig_xcounts}
\end{figure}

The resolved CXB is dominated by AGN populations which have been studied in detail by Chandra and XMM-Newton out to $z\sim 3$.  For X-ray AGN we use the population model of \citet{Gilli07}  based on X-ray luminosity functions and evolution of AGN. The models consider the observed XLFs, $k$-corrections, absorption distribution and spectral shapes of AGN and return the observed X-ray flux distribution at any redshift. The models have been shown to adequately reproduce source counts, redshift distribution and intrinsic column densities. Our adopted AGN population contains sources in the 0$<$z$<$8 range with a wide range in luminosity,  $38<\log(L_X/{\rm erg~ s^{-1}})<47$, to allow for very faint unresolved sources. Column densities are $20<\log(N_H/{\rm cm^2})<26$. The evolution of the AGN XLF in \citet{Gilli07} is modeled with and without an exponential decay at z$>$2.7 on top of the extrapolated evolution from lower redshift parametrization of \citet{Hasinger05}, i.e. $\phi (L,z)=\phi (L,z_0)10^{-0.43(z-z_0)}$ with $z_0$=2.7. We do not include the decrease at high-$z$ which results in a  CIB$\times$CXB signal within a factor of 2 of the case with a decline. In Figure \ref{fig_xcounts} we show the counts from our adopted AGN model and compare with data.

The extent to which AGN are removed by the joint IR/X-ray mask of C13 is estimated based on data from \citet{Civano12} who provide X/IR flux ratios for 1761 sources in the COSMOS survey reaching $S_{\rm 0.5-2 keV} = 1.9\times 10^{-16}$\erg. Towards faint fluxes AGN tend to become brighter at 3.6\mic\ deviating from the classic X/O=0. We fit a linear relation to the 3.6\mic\ vs 0.5-2 keV distribution, $m_{\rm IR} = -1.5\log{S_X}-1.7$, and extrapolate to the faint regime with a large Gaussian dispersion of $\sigma$=1.5 mag. We then apply $m_{\rm lim}=25$ to extract the selection $\eta(S)$ for the X-ray removal (see Eqn \ref{eqn_selection}). The scatter $\sigma$=1.5 mag is chosen such that the resulting shot noise $P_{SN}^{\rm X}$ in the X-ray power spectrum matches the data. The selection is shown in the top panel in Figure \ref{fig_xcounts} demostrates the extended tail of unresolved X-ray AGN caused by the wide dispersion in their IR flux \citep[see ][]{Civano12}. In Figure \ref{fig_dfdz} we show the unresolved AGN CXB production rate as function of the redshift. The bulk of the CXB flux from undetected AGN comes from z$\sim$1.

\subsection{Diffuse Hot Gas and WHIM} \label{sec_xgas}

In Section \ref{sec_xgal} we calculated the CXB contribution of hot gas heated within galaxies. However, diffuse gas in groups and filaments (including the WHIM) also contributes to the CXB and was found to dominate the unresolved 0.5-2 keV CXB fluctuations of \citep{Cappelluti12}. Scaling relations indicate that the X-ray masking of C13 removes galaxy clusters and groups down to $\log(M/ M_\odot)=(12.5-13.5)$ (i.e. $kT$$<$1.5 keV) \citep{Finoguenov07}. Thus only the low luminosity (low mass) and warm population of galaxy groups and filaments contributes to the unresolved CXB.

Since this class of objects is difficult to model analytically, we describe their properties using a set of mock maps from \citet{Roncarelli12}, who used a cosmological hydrodynamical simulation to define the expected X-ray surface brightness due to the large scale structures. The original hydrodynamical simulation (see the details in \citet{Tornatore10}) follows the evolution of a comoving volume of 37.5$h^{−1}$Mpc$^3$ considering gravity, hydrodynamics, radiative cooling and a set of physical processes connected with the baryonic component, among which a chemical enrichment recipe that allows to follow the evolution of seven different metal species of the intergalactic medium (IGM). From its outputs, \citet{Roncarelli12} simulated 20 lightcones with a size of $\sim$0.25 deg$^2$ each covering the redshift interval 0$<$z$<$1.5. Each pixel of the maps contains information about the expected observed spectrum in the 0.3--2.0 keV band with an energy resolution of 50 eV. The emission coming from the IGM was computed assuming an emission from an optically thin collisionally-ionized gas (Apec in XSPEC) model and considering the abundances of the different metal species provided by the simulation.

These maps/spectra have been convolved with the Chandra response in order to reproduce the effective Chandra count rates. Since the CXB data of C13 are masked for galaxy clusters, we apply a source masking on the simulated maps. We have simulated observations with the actual depth of the C13 field. We have added an artificial isotropic particle and cosmic background according to the levels estimated directly from the maps of C13. Random Poisson noise was artificially added to the image and we ran a simple wavelet detection with a signal to noise ratio threshold of 4. We have then excluded all the regions within which the overall encircled signal from sources is above 4$\sigma$ with respect to the background. The unresolved CXB production rate averaged over all the realizations is shown in Fig. \ref{fig_dfdz}.

\section{ The Angular Auto/Cross Power Spectrum of Multiple Populations } \label{sec_cross}

\begin{figure*}
\begin{center}
      \includegraphics[width=0.80\textwidth]{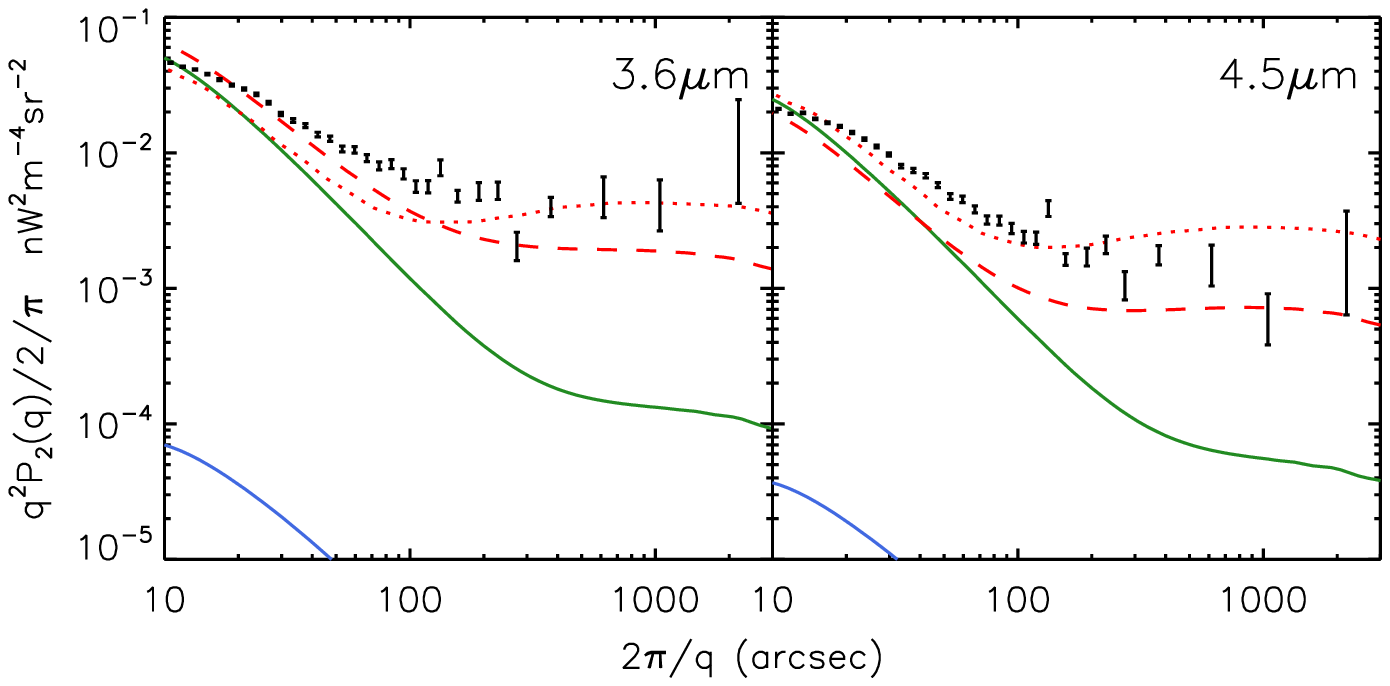}
      \caption{ The auto power spectra of the unresolved CIB fluctuations at 3.6\mic\ and 4.5\mic. Data points are from \citet{Kashlinsky12}. The contribution from known galaxies and AGN are show as green and blue line respectively (the AGN contribution barely visible above the plot range). The hypothetical IHL contribution is shown as red dashed line. We compare this with the original IHL model from \citet{Cooray12b} (dotted lines). All models are convolved with the IRAC beam taken from \citet{KAMM1,Arendt10}. }
    \label{fig_irauto}
\end{center}
\end{figure*}
\begin{figure}
      \includegraphics[width=0.45\textwidth]{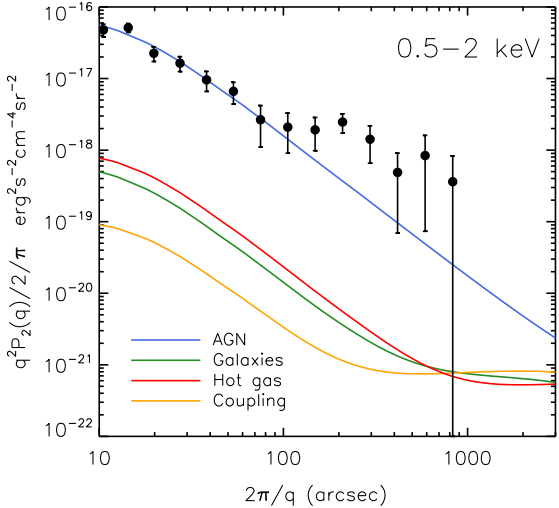}
      \caption{ The auto power spectra of the unresolved X-ray background fluctuations from different populations at the levels of C13. The contribution of normal galaxies (mostly XRBs) is shown in green and AGN in blue and hot/warm gas in red. We also display the net coupling term of the three components (orange). }
    \label{fig_xauto}
\end{figure}

Cosmic background fluctuations can be described by the angular power spectrum, $P(q)$. This can be written as the sum of the 
clustering and shot noise of the underlying source populations
\begin{equation}
  P_{\rm tot}(q)  = P(q) + P_{\rm SN}
\end{equation}
where $q$ is the angular wavenumber in rad$^{-1}$. The first term, representing the clustering, can be related to the three dimensional 
power spectrum of the underlying sources, $P(k,z)$, by projection via the Limber equation \citep{Limber53}
\begin{equation} \label{eqn_auto}
P(q) = \int \frac{H(z)}{cd_c^2(z)} \sum_i  \sum_{j\geq i} \left[ \frac{dF}{dz} \right]_i \left[ \frac{dF}{dz} \right]_j P_{ij}(qd_c^{-1},z) dz
\end{equation}
where $d_c(z)$ is the comoving distance and $H(z)$=$H_0\sqrt{\Omega_M(1+z)^3 + \Omega_\Lambda}$. The quantities in the square brackets are the unresolved flux production rates constructed in the previous sections for different source populations which are denoted by the indices $i$ and $j$ running over our three populations, 1: galaxies, 2: AGN, 3: diffuse emission. The summation results in six terms, three auto power terms ($i=j$) and three cross terms ($i\neq j$) that represent the coupling of different populations that live at the same epochs sharing the same environments. The shot noise arises from the fluctuation in the number of sources within the instrument beam. It only depends on the flux distribution of sources and can be expressed as
\begin{equation}
  P_{\rm SN} = \int_0^{S_{\rm lim}} S^2 \sum_i \left[ \frac{dN}{dS}\right]_idS
\end{equation}
where $S_{lim}$ is the minimum detected source brightness and $i$ denotes the source population as before. Note, that the shot noise, sometimes called Poisson term, does not have coupling terms as it represents a random process and is uncorrelated between different populations. Figures \ref{fig_irauto} and \ref{fig_xauto} show the modeled auto power spectrum of the angular fluctuations in the CIB and the CXB respectively, comparing them with current measurements.

In the description above, ``coupling'' terms ($i\neq j$) refer to the correlation of different populations at the same wavelength, not cross power between two wavelengths. The CIB$\times$CXB cross power spectrum can be written\footnote{Just as the two point correlation function is related to the auto power spectrum, the cross power spectrum is simply the Fourier transform of the cross-correlation function, $C_{nm}(\theta) = \langle \delta F_n(x)\delta F_m(x+\theta)\rangle$.}
\begin{equation} \label{eqn_cross}
  P_{\rm tot}^{\rm X,IR}(q) = \int  \frac{H(z)}{c d^2_c(z)} \sum_i\sum_j \left[ \frac{dF^{\rm X}}{dz}\right]_i \left[\frac{dF^{\rm IR}}{dz}\right]_j P_{ij}(qd_c^{-1},z)  dz
\end{equation}
where the summation results in nine terms representing all combinations of cross correlated X-ray and IR contributions from different source populations (note the different summation over $j$ compared to the conditional $j\geq i$ in Equation \ref{eqn_auto}). The cross power of the shot noise term is
\begin{equation} \label{eqn_sn}
  P_{\rm SN}^{\rm IR,X} = \int_0^{S^{\rm IR}_{\rm lim}}\int_0^ {S^{\rm X}_{\rm lim}} S^{\rm IR}S^{\rm X} \frac{d^2N}{dS^{\rm IR}dS^{\rm X}}dS^{\rm IR} dS^{\rm X}
\end{equation}
which is added to $P^{\rm IR,X}(q)$. This expression however, requires additional knowledge of the $dS_{\rm IR}/dS_{\rm X}$ dependence of each population in order to be evaluated. As our model construction lacks this information we use instead
\begin{equation}
  P_{\rm SN}^{\rm IR,X} = \int \left[ \frac{dP_{\rm SN}^{\rm IR} }{dz}\frac{dP_{\rm SN}^{\rm X}}{dz} \right]^{1/2}dz.
\end{equation}
We test the accuracy of this Equation using the Millennium lightcones which give $P_{\rm SN}^{\rm IR,X}$ directly and find it to be a good approximation to Equation \ref{eqn_sn} (see Figure \ref{fig_millcross}).

\subsection{Halo Model}

Our description of angular fluctuations requires knowledge of the power spectrum of luminous sources, $P_{ij}(k)$ (see Equations \ref{eqn_auto} and \ref{eqn_cross}). The distribution of sources inside the same collapsed dark matter halos can be related to the $\Lambda$CDM matter density field adopting a {\it Halo Occupation Distribution} (HOD) within a widely used halo model formalism \citep{CooraySheth02}. In this description, the power spectrum of clustering can be approximated as the sum of two terms
\begin{equation} \label{eqn_halo}
  P_{ij}(k) = P^{1h}_{ij}(k) + P^{2h}_{ij}(k),
\end{equation}
a one-halo term, $P^{1h}$, describing the correlated fluctuations between sources within the same parent halo, and a two-halo term, $P^{2h}$, arising from the spatial correlation of two sources hosted by separate parent halos. As before, the $i\neq j$ terms represent coupling terms between different populations, whereas for $i=j$ the expressions reduce to the more familiar form of \citet{CooraySheth02}. For source populations $i$ and $j$ these can be written
\begin{align} \label{eqn_onetwo}
  P^{1h}_{ij}(k) &= \int \frac{dn}{dM} \frac{(N_i^s N_j^c +  N_j^s N_i^c) u(k|M) + N_i^s N_j^s u^2(k|M)  }{\bar{n}_i\bar{n}_j }dM \\
  P^{2h}_{ij}(k) &= P^{\rm lin}(k) ( B_i^c + B_i^s)(B_j^c + B_j^s)
\end{align}
where
\begin{align} \label{eqn_B}
    B_i^c &= \int \frac{dn}{dM} \frac{N_i^c}{\bar{n}_i}b(M) dM \\
    B_i^s &= \int \frac{dn}{dM} \frac{N_i^s}{\bar{n}_i}b(M) u(k|M) dM.
\end{align}
where $i \neq j$ represent cross terms and $i=j$ reduces these expression to the more familiar form of \citet{CooraySheth02}. The individual quantities are defined as follows
\begin{itemize}
\item[-] $dn/dM$ is the evolving halo mass function for which we use the formalism of \citet{ShethTormen01}
\item[-] $\langle N_i^c(M,z) \rangle$ ($N_i^c$ shorthand) is the average halo occupation of central sources
\item[-] $\langle N_i^s(M,z) \rangle$ ($N_i^s$ shorthand) is the average halo occupation of satellite sources
\item[-] $\bar{n}_i(z)$ is the average number density of population $i$ such that
\begin{equation}
  \bar{n}_i = \int \left( \langle N_i^c(M,z) \rangle + \langle N_i^s(M,z) \rangle \right)\frac{dn}{dM} dM.
\end{equation}
\item[-] $P^{\rm lin}(k,z)$ is the linear $\Lambda$CDM power spectrum computed using the transfer function of \citet{EisensteinHu98} and the adopted cosmological parameters.
\item[-] $u(k|M)$ is the normalized Fourier transform of the NFW halo profile \citep{NFW96,CooraySheth02}
\item[-] $b(M,z)$ is the linear halo bias adopted from the ellipsoidal collapse formalism of \citet{ShethTormen01}
\end{itemize}
It should be kept in mind that all quantities in Equations (\ref{eqn_halo})--(\ref{eqn_B}) are evolving with redshift but we have omitted this dependence in the expressions to keep the notation tidy.

\subsection{Halo Occupation, Bias and Mass Selection} \label{sec_hod}

\begin{figure}
      \includegraphics[width=0.45\textwidth]{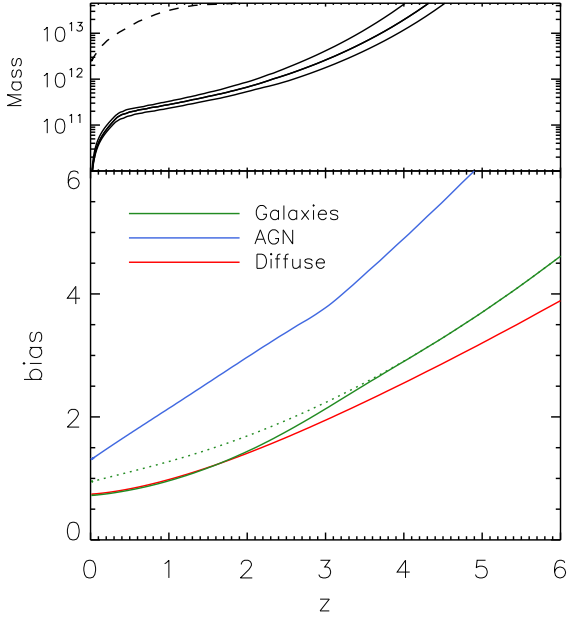}
      \caption{{\it Upper:} The approximate lowest halo mass removed by the IR mask as a function of redshift obtained from the Millennium SAM of \citet{Guo11}. The three lines correspond to 80,90,95\% of systems in the Millennium catalog being removed (top to bottom). The dashed line indicates the approximate removal threshold of groups/clusters in the X-ray maps \citep[see ][]{Erfanianfar13} {\it Lower:} The large scale bias, $b_{\rm eff}$ (Eqn. \ref{eqn_beff}), from galaxies, AGN and diffuse components. The solid lines show the biasing for a masked density field with the mass selection function in the upper panel whereas the dotted lines show the case of no source subtraction. Notice that at low redshift, the density field of undetected systems is underbiased. }
    \label{fig_bias}
\end{figure}
The halo occupation distribution (HOD) has been widely investigated for both galaxies and AGN. We adopt standard parameterizations of $N^c(M)$ and $N^s(M)$ which have been found to match observed data. For galaxies, we assume the four parameter description of \cite{Zheng05}\footnote{ $N_i^s$ is a shorthand notation for $\langle N_i^s(M,z)\rangle$ and is in general redshift dependent.}
\begin{align}\label{eqn_Ngal}
  N^c_{\rm gal}  &= \frac{1}{2}\left[ 1 + {\rm erf}\left( \frac{\log M - \log M_{\rm min} }{\sigma_{\log M}}\right) \right], \\
  N^s_{\rm gal}  &= \frac{1}{2}\left[ 1 + {\rm erf}\left( \frac{\log M - \log 2M_{\rm min} }{\sigma_{\log M}}\right) \right]  \left(\frac{M}{M_{sat}} \right)^{\alpha},
\end{align}
where $M_{\rm min}$ is the minimum halo mass that can host a central galaxy and $\sigma_{\log M}$ controls the width of the transition of the step from zero to one central galaxy. The satellite term has a cut-off mass which is twice as large as the one for central galaxies and grows as a power-law with a slope of $\alpha$ and is normalized by $M_{\rm sat}$. The amplification of the fluctuations through large scale biasing is most sensitive to the choice of $M_{\rm min}$ which is not well known as galaxies in the lowest mass halos are not detected. At a given redshift  however, the source subtraction removes the massive halos from top-down and limits the range of mass scales where the unresolved fluctuation signal arises, $M_{\rm min} < M < M_{\rm cut}(z)$, where $M_{\rm cut}(z)$ is the lowest mass halo removed at $z$ (see Figure \ref{fig_bias}, upper). We have adopted the following parameters of the HOD-model motivated by SDSS measurements of \citet{Zehavi11}: $\sigma_{\log M}=0.2$, $M_{\rm min} =10^{10}M_\odot$, $M_{\rm sat} = 5\times 10^{10}M_\odot$, and $\alpha = 1$ where we have deliberately chosen a low cutoff mass, $M_{\rm min}$, allowing low mass halos hosting galaxies well into the unresolved regime.

The HOD of AGN is less certain due to low number statistics but AGN seem to be preferentially found in halos of $\sim 10^{12.5}M_\odot$. The HOD has been measured for X-ray selected AGN at z$\lsim$1 \citep{Miyaji11,Allevato12,Richardson13} and optical quasars out to z$\sim$3 \citep{Shen13,Richardson12}. Previous studies suggest that, compared to optical quasars, X-ray selected AGN are more strongly clustered and reside in more massive host halos but the host halo mass range is insufficiently constrained for a definitive conclusion \citep{Nicoreview12}.  Like galaxies, the central AGN are modeled as a softened step function at $M_{\rm min}$ but the satellites are described by a power law with a low mass rolloff
\begin{align} \label{eqn_Nagn}
  N^c_{\rm AGN}  &= \frac{1}{2}\left[ 1 + {\rm erf}\left( \frac{\log M - \log M_{\rm min} }{\sigma_{\log M}}\right) \right], \\
  N^s_{\rm AGN}  &=  \left(\frac{M}{M_1} \right)^\alpha\exp{\left(-\frac{M_{\rm cut}}{M}\right)}
\end{align}
This description has five free parameters: $M_{\rm min}$, the characteristic mass scale of the step where the HOD goes from zero to a single AGN per halo, with the transition width controlled by $\sigma_{\log M}$. The mass at which a halo contains on average one satellite AGN is described by $M_1$; $\alpha$ is the power-law index controlling the steepness of the satellite HOD with increasing host mass; $M_{\rm cut}$ is the mass scale below which the satellite HOD decays exponentially. We have chosen parameters obtained in a numerical study of \citet{Chatterjee12} which agree with measured values when a selection of 
$L_{\rm bol}>10^{42}$erg/s is applied. We interpolate the redshift evolution of the $L_{\rm bol}>10^{42}$erg/s parameters given in Table 2 of \citet{Chatterjee12}.

The HOD of our diffuse component is somewhat uncertain as it does not describe the same population in X-rays and IR i.e. hot/warm gas as opposed to diffuse starlight. We allow the diffuse component to trace the NFW halo profile by considering a satellite term only, setting the HOD of central sources to zero. For diffuse IR, we adopt the parameters from the IHL model of \citet{Cooray12b} discussed in Section \ref{sec_ihl}, and assume that this component arises in halos in the $\sim 10^9-10^{12}M_\odot$ range. However, gas does not reach sufficient temperatures in such small halos but we neglect this by allowing hot gas is to live in halos anywhere below the mass limit of detected (and removed) groups and clusters identified by \citet{Erfanianfar13} in the EGS field (see dashed line in Fig. \ref{fig_bias}). 

When faced with source-subtracted images, the density field is modified in the process of masking the brightest sources which live in the most massive, and consequently, most biased halos. This effect can be accounted for by knowing the mass dependent luminosity distribution i.e. the conditional luminosity function. Since we do not have this information, we use the semi-analytical model of \citep{Guo11} mapped onto the Millennium Simulations to explore the halo mass dependence of source removal in deep IR maps. We eliminate all galaxies brighter than $m_{\rm lim}$=25 and construct a mass selection function defined as the fraction of unresolved systems as a function of host halo mass
\begin{equation}
  \eta(M,z) = \frac{N(M,z|>m_{\rm lim})}{N(M,z)}.
\end{equation}
We display this function in Figure \ref{fig_bias} (upper panel) which shows the mass scale at which 80\%, 90\% and 95\% of the systems are removed as a function of redshift. We multiply our galaxy HOD (Eqn. \ref{eqn_Ngal}) by this function, thereby subtracting the massive halos from the density field (including their satellites). Since this function is derived from galaxies, we do not apply this mass selection to the AGN and diffuse component and instead use the X-ray cluster/group detection limits of \citet{Erfanianfar13} as the upper mass limits. The overall effect of this is shown in Figure \ref{fig_bias}.

The amplification of fluctuations through large scale biasing of the sources follows from Equations \ref{eqn_B} in the limit where $u(k|M)\rightarrow 1$, or equivalently
\begin{equation} \label{eqn_beff}
  b^{\rm eff}_i(z) =  \int \frac{dn}{dM} \frac{N_i^c+N_i^s}{\bar{n}_i}b(M,z) dM
\end{equation}
where the mass dependent bias comes from the prescription of \citet{ShethTormen01}. The quantity is shown in Figure \ref{fig_bias} for galaxies, AGN and diffuse emission. We also show the bias without halo subtraction due to $\eta(M)$ i.e. in the absence of source masking. Note, how the density field becomes underbiased at low redshifts where the mask is most effective.

\subsection{Comparing fluctuation models with N-body simulations}

\begin{figure}
\begin{center}
      \includegraphics[width=0.49\textwidth]{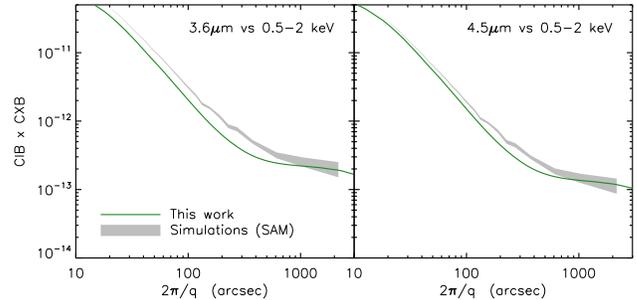}
      \caption{ The net cross-power spectrum $q^2P^{\rm IR,X}(q)/2/\pi$ from normal galaxies in units of ${\rm erg~ s^{-1}cm^{-2}nW m^{-2}sr^{-2}}$. The green line shows the prediction from our population model calculated using the Limber equation and the halo model formalism described in the text. The gray areas show the result from directly Fourier transforming simulated images, $P_{\rm IR,X}(q) = \langle \Delta_{\rm IR}(q)\Delta_{\rm X}^*(q) \rangle$, obtained from a semi-analytic models based on the Millennium Simulation \citep{Guo11}. }
    \label{fig_millcross}
\end{center}
\end{figure}
\begin{figure*}
      \includegraphics[width=.98\textwidth]{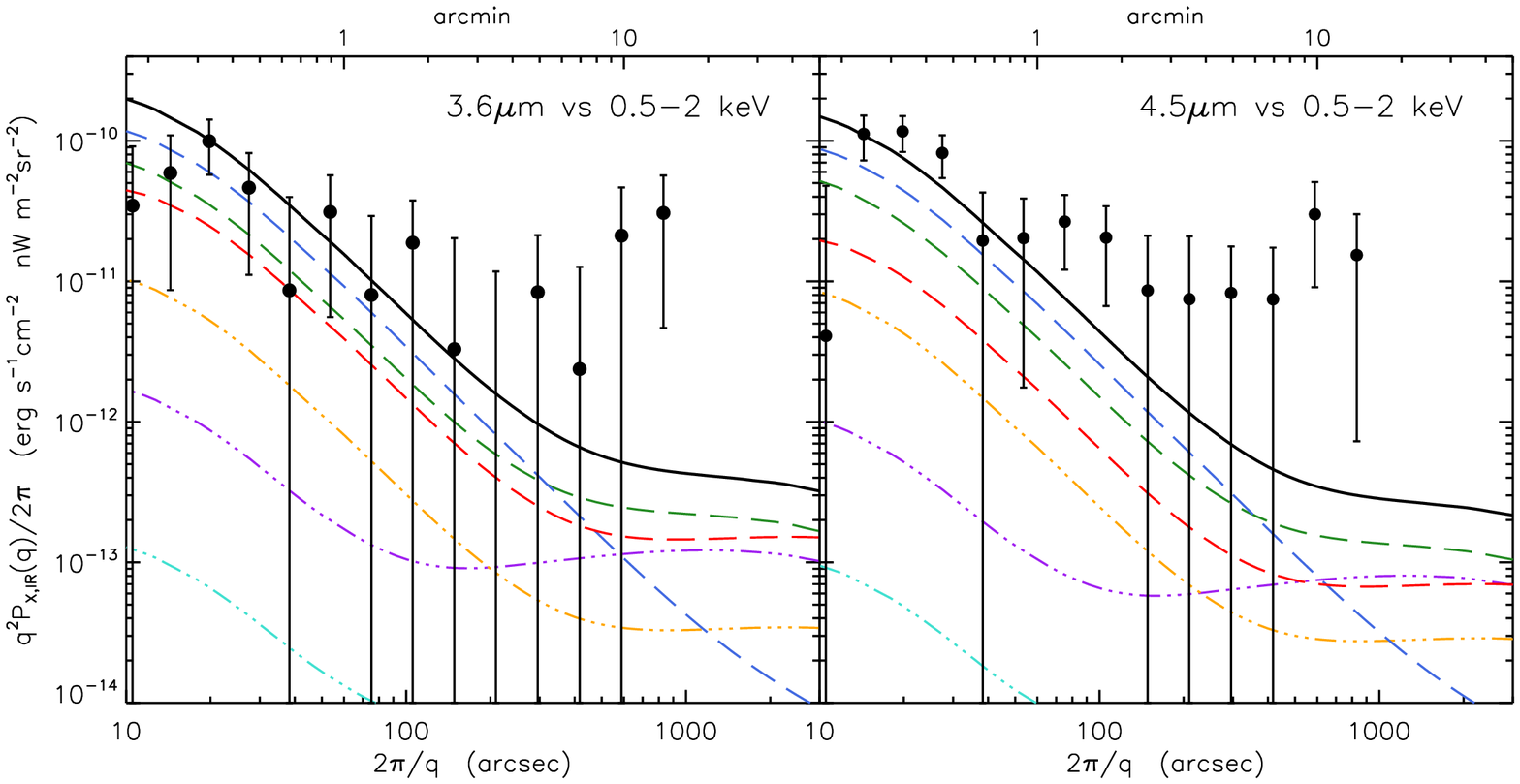}
      \caption{ The X-ray versus IR cross-power terms of all source populations compared to measurements from \citet{Cappelluti13}. The total cross-power spectrum from all terms is shown as solid black line. The individual terms are galxies$_{\rm IR}$-galaxies$_{\rm X}$ (green), galaxies$_{\rm IR}$-AGN$_{\rm X}$ (purple), galaxies$_{\rm IR}$-diffuse$_{\rm X}$ (orange), AGN$_{\rm IR}$-galaxies$_{\rm X}$ (turquoise), AGN$_{\rm IR}$-AGN$_{\rm X}$ (blue) and diffuse$_{\rm IR}$-diffuse$_{\rm X}$ (red). Auto terms are shown as dashed lines ($i=j$) whereas coupling terms ($i\neq j$) are shown as dashed-dotted lines. }
      \label{fig_cross}
\end{figure*}
In real measurements, the angular power spectrum is obtained directly from the masked and Fourier transformed image, $P_{mn}(q) = \langle \Delta_m(q)\Delta_n^*(q) \rangle$. However, our calculation of $P(q)$ relies on the projection of emitting populations via the Limber equation with empirically motivated assumptions for their HOD. A more sophisticated treatment would link the source luminosities to host halo masses in a conditional luminosity function, $\Phi(L(M))$ \citep[see e.g.][]{Cooray06,Bethermin13}. In order to test the validity of the approximations made, we make use of the Millennium Simulation SAM of \citet{Guo11} to derive the unresolved CIB$\times$CXB power spectrum from galaxy populations. The evolving simulation box has been projected to construct lightcones that provide 2 deg$^2$ mock images of the extragalactic sky based on the SAM \citep{Henriques12}. From the mock catalogs, we remove all galaxies brighter than IRAC$_{3.6,4.5}$ magnitude 25 AB including substructure associated with the parent halo. The X-ray emission is calculated using the relation of \citet{Lehmer10} $L_X = \alpha M^\star + \beta {\rm SFR} + (1+z)^{0.5}$ where we have added the last term to account for evolution (see Section \ref{sec_xgal}).

We calculate the source-subtracted fluctuations directly from the mock images, $\langle \Delta_m(q)\Delta_n^*(q) \rangle$, and compare the results with our fluctuation model in Figure \ref{fig_millcross}. Despite the difference in approach, this SAM and our population model predict consistent fluxes and CIB$\times$CXB cross-power. Caveats worth mentioning include the halo resolution limit of Millennium $\sim10^{10}M_\odot$ and the possibility of spurious power arising from the replication of the simulation box required for the lightcone construction \citep[see ][]{Blaizot05}. The agreement with our models is nevertheless encouraging.

\section{Results} \label{sec_results}

\subsection{ CIB Fluctuations }

Figure \ref{fig_irauto} shows the auto power spectrum of CIB fluctuations from unresolved galaxies, AGN and IHL, comparing them with the measurements of C13. The contribution from unresolved galaxies (green) is discussed in detail in \citet{Helgason12} and the IHL (red) is close to that of \citet{Cooray12b}. We only use the default model from \citet{Helgason12} which has been validated in the SEDS survey \citep[Fig. 34 in][]{Ashby13} reducing the uncertainties of the faint-end extrapolation of the luminosity function. The contribution from AGN (blue) is much smaller due to their low numbers compared to IR galaxies, $<10\%$. In addition, C13 found that their source-subtracted CIB power spectrum is independent of the X-ray mask. This means that X-ray flux that may be missed by the IR mask, such as from the wide wings of the extended Chandra PSF, will show up in the CXB power spectrum but will not contribute to the cross-power CIB$\times$CXB.

\subsection{ CXB Fluctuations }

We find that the CXB power spectrum is dominated by shot noise from unresolved AGN (see Fig \ref{fig_xauto}) with the contribution from galaxies and gas being considerably lower. However, a different study \citet{Cappelluti12} finds the largest contribution to come from hot gas. There are various reasons for the different findings. First, the X-ray maps of \citet{Cappelluti12} are deeper (4Ms) than C13 (800ks) allowing \citet{Cappelluti12} to directly mask AGN to much fainter levels. Second, \citet{Cappelluti12} modeled AGN in the luminosity range 42$<$$\log{(L/L_\odot)}$$<$47 whereas we include AGN all the way down to $\log{(L/L_\odot)}$=38. This makes a substantial difference in the abundance of faint AGN.

The net CXB from unresolved galaxies and AGN is $2.1\times 10^{-13}$ and $7.9\times 10^{-14}\ergdeg$ respectively. Whereas the CXB power spectrum is consistent with being entirely due to shot noise from unresolved AGN there is a hint of additional clustering towards large scales which is not accounted for. No reasonable amount of clustering (bias $\lsim$30) is sufficient to account for enhanced CXB fluctuations on scales $>$200$^{\prime\prime}$. If this component is real and extragalactic in nature, it could indicate the source of the coherence with the CIB fluctuations. However, it is important to note that the 0.5--2 keV fluctuations are contaminated by foreground emission from the Galaxy which is not sensitive to the removal of extragalactic point sources. Any interpretation of the CXB power spectrum therefore carries an intrinsic source of uncertainty due to the contribution of the Galaxy. A non-negligible Galaxy component at $<$1 keV could explain why C13 measure a low-level of cross-correlation between [0.5-2] keV and [2-4.5]--[4.5-7] keV maps. While irrelevant for the CXB$\times$CIB cross-power spectrum, correcting for the Galaxy would reduce the measured CXB power spectrum. Additionally, the extended point-spread function (PSF) of Chandra could spread some fraction of the X-ray point source flux outside the finer IR mask. This would not show up in the CIB$\times$CXB cross-power as the large scale CIB fluctuations do not correlate with either IR or X-ray removed sources.

\subsection{ CIB$\times$CXB Fluctuations }

We start with summarizing the resultant contributions to the cross-power from three main components appearing in Figure \ref{fig_cross}: galaxies, AGN and diffuse:
\begin{figure}[b]
      \includegraphics[width=0.48\textwidth]{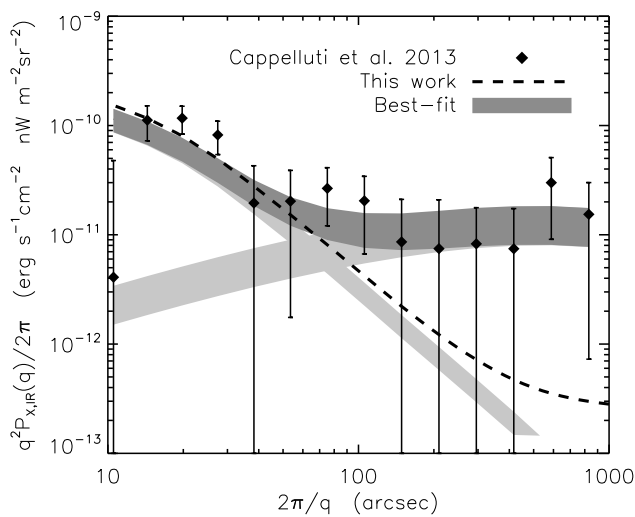}
      \caption{ The 4.5 vs 0.5--2 keV cross power spectrum. Data points are from C13. The dark gray region corresponds to the 1$\sigma$ uncertainty in the best fit model $P(q) = a_1P_{\rm \Lambda CDM}+a_2$ with the two individual terms shown as light gray regions. Our model of net contribution from galaxies, AGN and diffuse emissions is shown as black dashed line. }
    \label{fig_regress}
\end{figure}
\begin{itemize}
\item {\bf Galaxies}. The largest contributions to the cross-power comes from galaxies (${\rm gal_{IR}\times gal_X}$, green line) and AGN (${\rm gal_{IR}-AGN_X}$, purple line). A significant galaxy--galaxy component is expected because i) they make-up a substantial unresolved CIB component and ii) dominate AGN at faint X-ray fluxes (see Fig. \ref{fig_xcounts}). The small scale cross-power is in good agreement with the data, the ${\rm gal_{IR}\times gal_X}$ shot noise making up $\sim$30\%. This fraction decreases towards large scales however. In order to explain the shape of the CIB$\times$CXB fluctuations at all scales in terms of galaxies only, one needs to alleviate the problem of the low clustering with respect to shot noise, as is the case with the CIB fluctuations. Simply increasing the flux of the unresolved populations would overproduce the small scale power. A way of increasing the large scale power without affecting the shot noise is to enhance the galaxy bias. However, this requires bias of $\gsim$10 which is not expected for faint low-mass systems. We therefore conclude that the entire CIB$\times$CXB signal cannot originate from normal galaxies unless future measurements show that the cross-fluctuations ($q^2P(q)/2/\pi$) decrease towards large scales as opposed to staying roughly flat as indicated by the C13 data.

\item {\bf AGN}. Shot noise from unresolved AGN provides the largest contribution, $\sim$60\% to the small scale CIB$\times$CXB power (Fig \ref{fig_cross}, blue line). This is because a greater fraction of bright AGN remains unresolved after IR masking (see Figure \ref{fig_xcounts}). At large scales, their AGN$_{\rm IR}\times$AGN$_{\rm X}$ contribution is small due to lower flux in both IR and X-rays compared with galaxies. However, a substantial contribution comes from X-ray AGN correlating with IR galaxies (purple line). As shot noise does not appear in this term, the cross power spectrum has a shape that resembles the data but with an amplitude which is more than an order of magnitude below the data. To test whether this term could be enhanced, we examined the case in which the IR source subtraction removes no additional AGN i.e. only AGN detected in X-rays are removed. This gives an amplitude that is one order of magnitude below the data points. Enhancing the clustering of the AGN population to the levels of very biased high-$z$ quasars, corresponding to AGN living in $>10^{13}M_\odot$ halos, still falls below the measured levels. In fact, both our AGN population model \citep{Gilli07} and our AGN removal selection \citep{Civano12} are chosen conservatively and should, if anything, give a smaller signal.

\item {\bf Diffuse}. Dispersed starlight around and between masked galaxies can share the same environment with diffuse warm gas in collapsed halos and filaments. For distant structures however, the thermal spectrum of the ionized gas, $\sim$1 keV, quickly redshifts out of the 0.5-2 keV band (see Section \ref{sec_xgas}) and has therefore limited correlations with the IHL which mostly arises at different epochs, $z\sim 1-4$. Despite this, the large scale CIB$\times$CXB component arising between IHL and warm gas at $z<1$ is comparable to that of galaxies and AGN (see Figure \ref{fig_cross}, red). The diffuse component could be made larger if the bulk of IHL emission arose at low-$z$ in halos $\gsim 10^{12.5}M_\odot$, but this would be unlikely to explain the entire CIB$\times$CXB data. Coherence between IHL and X-ray galaxies/AGN is also difficult to accommodate for the following reasons. In the resolved regime, the point sources are masked so any X-ray emission originating in the central regions is eliminated with no correlation with diffuse IR light outside the mask. In the unresolved regime, the IR galaxies themselves should dominate over IHL which can only constitute a fraction of the total galaxy light. We already account for the coherence of unresolved IR galaxies with X-ray galaxies/AGN. Furthermore, the fact that the CIB fluctuations are not sensitive to the X-ray mask argues against the removed X-ray sources being responsible for the CIB$\times$CXB signal. We note the additional problems with the IHL hypothesis in Sec. \ref{sec_ihl}.

\end{itemize}

\begin{table} 
\begin{center}
\caption{Comparison of our net model with the best-fit model $P(q) = a_1 P_{\rm \Lambda CDM} + a_2$. The power is in units of ${\rm erg~ s^{-1}cm^{-2}nW~ m^{-2}sr^{-1}}$ }
\begin{center}
\begin{tabular}{ l c c c c }
\hline\hline

 & \multicolumn{2}{c}{ Clustering$^1$ } & \multicolumn{2}{c}{ Shot noise} \\
 & \multicolumn{2}{c}{ ($a_1 \times 10^{17}$) } & \multicolumn{2}{c}{ ($a_2 \times 10^{19}$)  } \\
& 3.6\mic\ &  4.5\mic\  &  3.6\mic\ &  4.5\mic\ \\
\hline
Best-fit & 2.5$\pm$2.1 & 4.3$\pm$1.7 & 1.0$\pm$0.4 & 1.3$\pm$0.3 \\
This work & 0.21 & 0.11 & 1.93 & 1.44 \\
\hline
\end{tabular}\end{center}
\tablecomments{ $^1$clustering at 1000$^{\prime\prime}$ }
\hfill
\label{tab_regress}
\end{center}
\end{table}

\begin{figure}[t]
\begin{center}
      \includegraphics[width=0.49\textwidth]{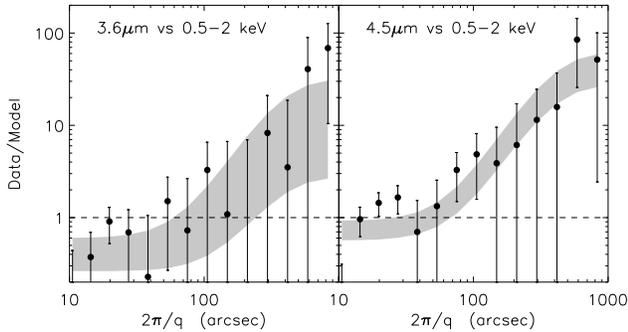}
      \caption{ The systematic discrepancy of the cross power spectra of our model and the data as a function of angular scale, $P_{\rm data}(q)/P_{\rm model}(q)$. The normalized model is indicated by the dashed line. The 1$\sigma$ regions of the best-fit two parameter model is show in light gray. }
    \label{fig_discrep}
\end{center}
\end{figure}

Despite the large uncertainties in the data, there seems to be a systematic lack of cross-power at the large scales $>300^{\prime\prime}$ where the source clustering is in the linear regime. At small scales, our modeled shot noise term is in agreement with the data. To better understand these results, we consider a simple model composed of linear $\Lambda$CDM clustering and a noise term, $P(q) = a_1 P_{\rm \Lambda CDM} + a_2$ where $a_1$ and $a_2$ are free parameters and $P_{\rm \Lambda CDM}$ is normalized to unity at 1000$^{\prime\prime}$. This model is then convolved with the Chandra response function. For 4.5\mic\ vs [0.5--2 keV], we find best-fit parameters $a_1=(4.3\pm 1.7)\times 10^{-17}$ and $a_2 = 1.3 \pm 0.3 \times 10^{-19}$ in ${\rm erg~ s^{-1} cm^{-2}nW~ m^{-2}sr^{-1}}$, resulting in a $\chi^2/12 = 1.2$ (see Figure \ref{fig_regress})\footnote{The smallest scale data point at $10.5^{\prime\prime}$ is offset with respect to the rest and as a result leads to smaller $a_2$ and drives up the $\chi^2$. This is why the gray best-fit region seems somewhat below the small scale data points. If we neglect this data point the best-fit model is brought in perfect agreement with our modeled shot noise.}. The net power from our model of $z<6$ sources (black line) is a poor fit to the data ($\chi^2/8=2.8$) falling more than an order of magnitude below the best-fit model. This distinction is not significant at 3.6\mic\ vs [0.5--2 keV] due to the large uncertainties but the systematically growing discrepancy towards large scales suggests the same behavior as 4.5\mic\ vs [0.5--2 keV]. This is illustrated in Figure \ref{fig_discrep}.

\section{Conclusions and Discussion} \label{sec_discussion}

In this paper we have considered known populations of X-ray sources at $z<6$ in an attempt to explain the measured spatial coherence of the unresolved CIB and CXB. The dominant contribution to the modeled CIB$\times$CXB signal comes from unresolved galaxies containing X-ray binaries and IR galaxies associated with X-ray emitting AGN found within the same large scale structures. However, we find that the combined contribution from galaxies, AGN and hot gas, is unable to produce the large scale cross-power needed to explain the data. At 4.5\mic\ vs 0.5--2 keV, the large scale cross-power is only $(2.6\pm 1.0)\%$ of the best-fit model, but the discrepancy decreases towards small angular scales where the shot noise becomes consistent with the best-fit. At 3.6\mic\ vs 0.5--2 keV these identifications are not robust due to larger uncertainties in the data. 

Warm gas in $\lsim {\rm 10^{13.5}M_\odot}$ systems can be bright at $z\lsim 0.5$ but IR sources are eliminated down to $\sim{\rm 10^{11}M_\odot}$ in this redshift range. A significant source of coherence may arise at low-$z$ in association with diffuse starlight (IHL). However, the correlation of IHL with unresolved X-ray galaxies/AGN at higher redshifts is problematic. Whereas at $z=0$ the IHL corresponds to $\sim$1\% of the total CIB produced by galaxy populations, at $z\gsim 2$ it seems to {\it exceed the total CIB from all galaxies} (see Figure \ref{fig_dfdz}). The lack of correlation between the source-subtracted CIB fluctuations and i) deep Hubble/ASC maps \citep{KAMM4} and ii) test halo images \citep{Arendt10}, also makes it difficult to favor missing starlight as a dominant component of the fluctuations.

While other mechanisms capable of producing a correlation between X-rays and IR may exist, they are generally expected to be much weaker than the dominant forms considered in this work: galaxies, AGN and hot gas. Thermal emission from hot dust $\sim$700K would have to arise in the local universe as it would otherwise redshift out of the near-IR bands and it is also inconsistent with the observed blue color of the source-subtracted CIB fluctuation in the 2.4-4.5\mic\ range. Because of its red colors, any dust dominated component would have to be underdominant in the CIB fluctuations while being associated with X-ray emission from the dominant CXB component i.e. the Galaxy foreground or obscured AGN. Infrared cirrus emission in the Galaxy should absorb X-rays and exhibit a {\it negative } cross-power contrary to the measurements of C13. In the case of obscured AGN, they make up a greater fraction of the hard CXB and are less significant in soft X-rays. This is not consistent with the fact that C13 detect CIB$\times$CXB cross-correlation in the 0.5--2 keV band but not in the harder bands. Furthermore, the cross-correlation between the [0.5--2] keV band and both [2--4.5] and [4.5--7] keV are small. There is a hint of clustering in the CXB power spectrum at $>$100$^{\prime\prime}$ which may or may not be the source of coherence with the CIB. However, the component producing the large scale CIB$\times$CXB cannot constitute less than $\sim$15--20\% of the CXB clustering.

The possibility that the CIB$\times$CXB signal is contributed by high-$z$ miniquasars is discussed in C13. Such objects are expected to form early and grow rapidly in order to explain the population of bright quasars already in place at $z\sim 6$. \citet{Yue13b} have constructed a population model of highly obscured direct collapse black holes that is able to account for 1) net CIB measurements and  $\gamma$-ray absorption constraints, 2) the shape and amplitude of the source-subtracted CIB fluctuations, 3) the unresolved soft CXB level, and 4) the shape and amplitude of the spatial coherence in the unresolved CIB$\times$CXB. The inclusion of such a component improves the CIB$\times$CXB best-fit considerably (see Fig \ref{fig_regress}). Whether these requirements can be realistically satisfied physically by other types high-$z$ miniquasars and at the same time stay within limits imposed by reionization and black hole mass growth, will be investigated in future work.

\section*{ACKNOWLEDGMENTS}

KH acknowledges useful discussions with R. Arendt, A. Ferrara, B. Lehmer and R. Mushotzky. This work was supported by NASA Headquarters under the NASA Earth and Space Sciences Fellowship Program Grant -- NNX11AO05H.


\label{lastpage}
\end{document}